\documentclass[fleqn,usenatbib]{mnras}

\usepackage{newtxtext,newtxmath}
\usepackage{longtable}
\usepackage{booktabs}
\usepackage[T1]{fontenc}
\usepackage{float}
\DeclareRobustCommand{\VAN}[3]{#2}
\let\VANthebibliography\thebibliography
\def\thebibliography{\DeclareRobustCommand{\VAN}[3]{##3}\VANthebibliography}

\usepackage{graphicx}	
\usepackage{amsmath}	

\title[80 pulsars from MWA incoherent drift scans]{A study of 80 known pulsars at 185 MHz using MWA incoherent drift-scan observations}

\author[Ting Yu et al.]{
Ting Yu,$^{1,2}$\thanks{E-mail: yuting@shao.ac.cn (SHAO)}
Hongyu Gong,$^{4}$
Zhifu Gao,$^{5}$
Zhongli Zhang,$^{1,3}$
Zhigang Wen,$^{5}$
Yujie Wang,$^{1,2}$
and Tao An$^{1,2,3}$
\\
$^{1}$Shanghai Astronomical Observatory, Chinese Academy of Sciences, Shanghai 200030, China\\
$^{2}$School of Astronomy and Space Science, University of the Chinese Academy of Sciences, Beijing 100012, China\\
$^{3}$State Key Laboratory of Radio Astronomy and Technology, A20 Datun Road, Chaoyang District, Beijing, 100101, P. R. China\\
$^{4}$Institute of Astronomy and Information, Dali University, Dali 671003, China\\
$^{5}$Xinjiang Astronomical Observatory, Chinese Academy of Sciences, Urumqi, Xinjiang 830011, China
}

\date{Accepted XXX. Received YYY; in original form ZZZ}

\pubyear{\the\year{}}

\begin{document}
\label{firstpage}
\pagerange{\pageref{firstpage}--\pageref{lastpage}}
\maketitle

\begin{abstract}
A systematic study of 80 known pulsars observed at 185 MHz has been conducted using archival incoherent-sum data from the Murchison Widefield Array (MWA). The dataset comprises 48 drift-scan observations from the MWA Voltage Capture System, covering $\sim$30,000 deg$^2$ of sky with sensitivities reaching $\sim$8 mJy in the deepest regions. An optimized \textsc{PRESTO}-based search pipeline was deployed on the China SKA Regional Centre infrastructure. This enabled the detection of 80 known pulsars—representing a $\sim$60\% increase over the previous census. Notably, this includes 30 pulsars with first-time detections at this frequency, of which pulse profiles and flux densities are presented. Spectral, scattering, and pulse-width properties were examined for the sample, providing observational constraints on low-frequency turnover, propagation effects, and width–period relations. This study highlights the value of wide-field, low-frequency time-domain surveys for constraining pulsar emission and propagation, offering empirical insights that may inform future observations with instruments such as SKA-Low.
\end{abstract}

\begin{keywords}
pulsars: general --- pulsars: searching pipeline --- pulsars: search --- pulsars:  catalogs
\end{keywords}



\section{Introduction}
\label{sec:intro}

Pulsars, first discovered through low-frequency observations in 1967 \citep{Hewish1968}, are highly magnetized, rotating neutron stars. They serve as unique laboratories for studying fundamental physics, stellar evolution \citep{Shapiro1983}, and the interstellar medium (ISM) \citep{Taylor1993, Cordes2002, Yao2017}. Early low-frequency studies revealed important propagation effects such as dispersion and scattering \citep{Rickett1977, Blandford1985}, as well as spectral turnovers in many sources \citep{Ochelkov1984}. Despite these early insights, the majority of confirmed pulsars have been discovered at observing frequencies above 400~MHz, with only $\sim20\%$ of the 4343 pulsars in the ATNF catalog\footnote{Catalog version 2.6.3, \url{http://www.atnf.csiro.au/research/pulsar/psrcat/} \citep{Manchester2005}} having detections below 350~MHz.

In recent years, significant progress has been made in low-frequency pulsar searches, facilitated by both traditional single-dish surveys and new-generation interferometric arrays. Notable efforts include the Arecibo All-Sky 327 MHz Drift Pulsar Survey (AO327) at 327~MHz \citep{Deneva2013, Deneva2016, Martinez2019}, the Green Bank North Celestial Cap Survey (GBNCC) at 350~MHz \citep{Stovall2014, Lynch2018, McEwen2020, Lynch2021}, and the Giant Metrewave Radio Telescope High-Resolution Southern Sky Survey (GHRSS) at 322~MHz \citep{Bhattacharyya2016, Bhattacharyya2019, Singh2022, Sunder2023, Singh2023}. Extending to even lower frequencies, coherent beamforming with LOFAR has yielded 76 discoveries from the LOFAR Tied-Array All-Sky Survey (LOTAAS) at 135~MHz \citep{Tan2018, Sanidas2019, Michilli2020, Tan2020, Wateren2023}, while the Pushchino Multibeam Pulsar Search (PUMPS) has identified new pulsars at 111~MHz \citep{Tyul'bashev2022, Tyul'bashev2024}. The Murchison Widefield Array (MWA), operating below 300~MHz, has also been employed for time-domain pulsar studies \citep{Xue2017, Bhat2023a, Bhat2023b}, leveraging its large field of view and flexible observing modes. 

High-dispersion measure (DM) pulsars are particularly difficult to detect at low radio frequencies, as interstellar dispersion and scattering broaden the pulses and reduce the sensitivity of conventional time-domain searches. Recent studies have shown that radio imaging provides a valuable complementary approach, being less affected by such propagation effects and enabling the discovery of ultra-long-period pulsars and transients. Notable examples include GPM J1839–10 \citep[22 minutes;][]{Hurley-Walker2023} and GLEAM-X J0704–37 \citep[2.9 hours;][]{Hurley-Walker2024}, with Galactic plane imaging further revealing additional candidates \citep{Mantovanini2025}. Theoretical work also suggests that some ultra-long-period sources may represent exotic compact objects \citep{Zhou2025}, underscoring the role of wide-field low-frequency imaging in identifying populations missed by traditional searches.

Initial pulsar observations with the MWA used incoherent beamforming to conduct a census of known sources \citep{Xue2017}, primarily targeting short-period pulsars ($P < 2$~s) and utilizing only the central regions of the MWA primary beam. Subsequent surveys adopted coherent-sum modes to improve sensitivity, albeit with limited sky coverage. One persistent challenge for all low-frequency searches is signal degradation from interstellar scattering and dispersion. As next-generation facilities such as the Square Kilometre Array \citep[SKA;][]{Dewdney2009} plan to survey the low-frequency sky \citep[50-350 MHz;][]{Keane2015}, characterizing these propagation effects and refining search techniques remain priorities.

Archival MWA Phase I Voltage Capture System data were reprocessed using a harmonically sensitive, FFT-based search pipeline \citep{Wei2023}, yielding a sky coverage of $\sim$30,000 deg$^2$. The dataset includes 80 detections of known pulsars, of which 30 represent first-time measurements of pulse profiles and flux densities at 185 MHz. This significantly expands the sample of pulsars with constrained low-frequency spectral and scattering properties, enabling refined characterization of spectral turnovers, scattering indices, and pulse width–period relations at metre wavelengths.

The paper is structured as follows. Section~\ref{sec:obs_reduction} describes the observations and data processing methods. Section~\ref{sec:results} presents the search outcomes and detailed analyses of spectral, scattering, and pulse width properties. In Section~\ref{sec:discussion}, we discuss implications for pulsar population studies and search methodologies. Finally, Section~\ref{sec:conclusions} summarizes our main conclusions.

\section{Archival data reduction}
\label{sec:obs_reduction}

This study analyzes 48 archival drift-scan observations from MWA Phase I \citep{Tingay2013, Bowman2013}, recorded with the Voltage Capture System \citep[VCS;][]{Tremblay2015} at 185 MHz with 30.72 MHz bandwidth. The VCS provides raw voltages at 100~$\mu$s and 10 kHz resolutions. The dataset corresponds to 45 hours of total on-sky integration, with individual scans ranging from 7 minutes to 1.4 hours, amounting to $\sim$ 1.3 PB. Each pointing covers $\sim$450 deg$^2$ in incoherent-sum mode. While partially overlapping with the sample of \citet{Xue2017}, the present set includes $\sim$25\% more observations and employs distinct processing and search strategies. Observation metadata are summarized in Appendix Table~\ref{appendix:obs_table}.

\begin{figure}
   \centering
  \includegraphics[width=8cm, angle=0]{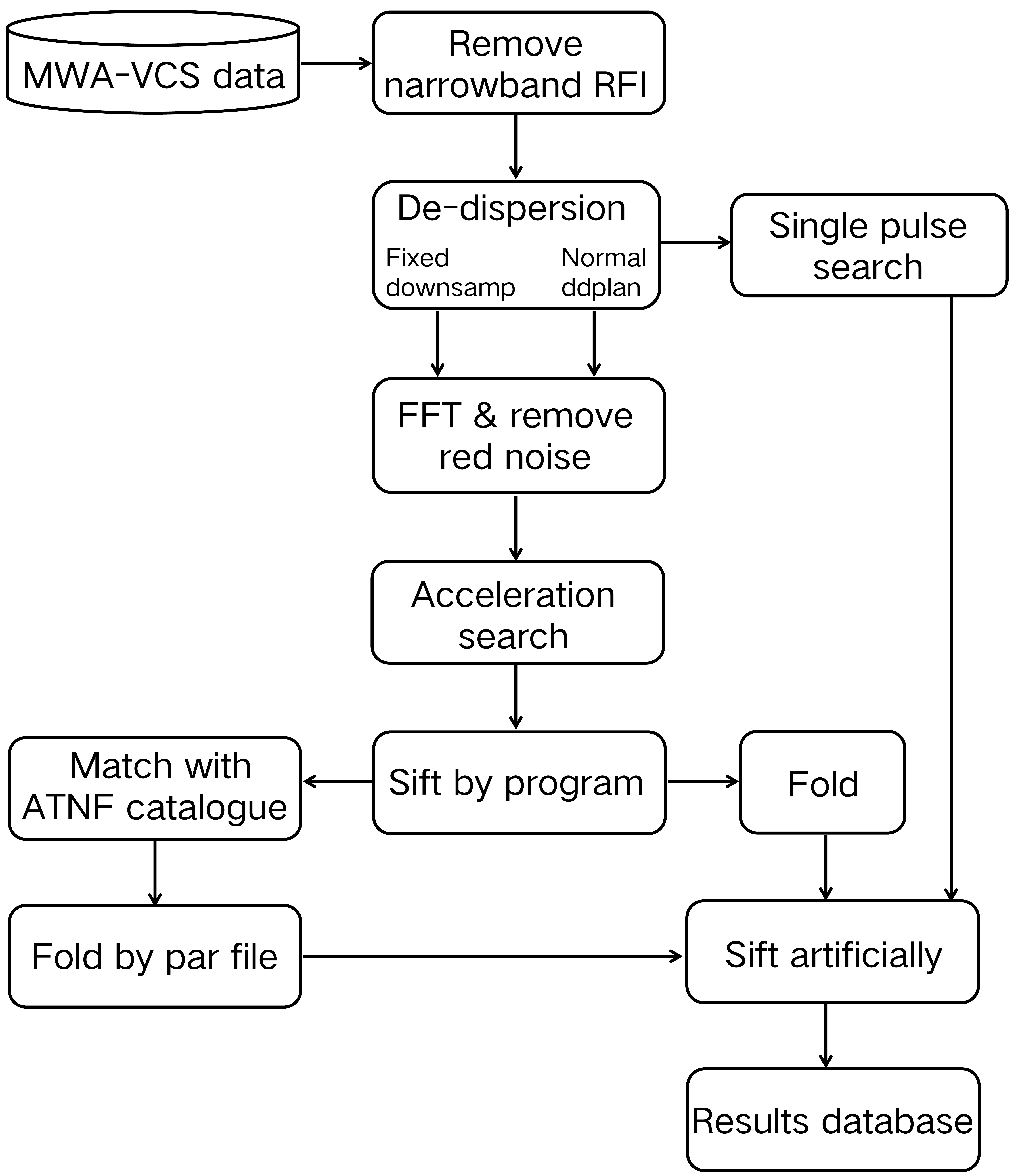}
   \caption{Pulsar search pipeline for the MWA-VCS incoherent-sum data.} 
   \label{fig:pipeline}
\end{figure}

The processing pipeline (Figure~\ref{fig:pipeline}) was implemented with \textsc{PRESTO} \citep{Ransom2001, Ransom2002}\footnote{https://www.cv.nrao.edu/~sransom/presto/} on the China SKA Regional Centre cluster \citep{An2019, An2022, Wei2023}. Radio Frequency Interference (RFI) was excised using \texttt{rfifind} (12-s intervals) to identify narrow- and broadband interference, with periodic signals removed via zero-DM filtering. Data were de-dispersed with \texttt{prepsubband}, adopting a stepped DM plan of 2555 trials (Table~\ref{tab:ddplan}) and fixed downsampling factors of 4, 8, and 16, in total four parallel approaches. The resulting time series were Fourier transformed with \texttt{realfft}, and low-frequency red noise was mitigated with \texttt{rednoise}. Periodicity searches employed \texttt{accelsearch} (16 harmonics, $z_{\max}=0$), with candidates sifted using \texttt{ACCEL\_sift.py}, folded using \texttt{prepfold}, and then visually inspected. A parallel single-pulse search was also performed with \texttt{single\_pulse\_search.py} to identify transient and long-period sources. On average $\sim$100 candidates per observation exceeded a 4$\sigma$ threshold, and over 20,000 were folded without exploring DM, $P$, or $\dot{P}$.

\begin{table}
\begin{center}
\begin{minipage}[]{12cm}
\caption{De-dispersion plan adopted in the present search} \label{tab:ddplan}
\end{minipage}
\setlength{\tabcolsep}{6pt}
\small
 \begin{tabular}{ccccccc}
  \hline\noalign{\smallskip}
$\mathrm{DM}_{\min}$ & $\mathrm{DM}_{\max}$ & $\delta \mathrm{DM}$ & $N_{\mathrm{DM}}$ & $d_{s}$ & $\Delta t_{\mathrm{eff}}$ & $n_{\mathrm{sub}}$ \\
 (pc cm$^{-3}$) & (pc cm$^{-3}$) & (pc cm$^{-3}$) & & & (ms) & \\
        \hline\noalign{\smallskip}
        1.0 & 22.9 & 0.02 & 1093 & 1 & 0.1 & 4  \\
        22.9 & 45.7 & 0.05 & 457 & 2 & 0.2 & 8 \\
        45.7 & 91.5 & 0.11 & 415 & 4 & 0.4 & 16 \\
        91.5 & 183.0 & 0.21 & 435 & 8 & 0.8 & 32  \\
        183.0 & 250.0 & 0.43 & 155 & 16 & 1.6 & 64  \\

  \noalign{\smallskip}\hline
\end{tabular}
\end{center}
\textbf{Note.} Columns 1–2 list the dispersion measure range 
($\mathrm{DM}_{\min}, \mathrm{DM}_{\max}$); 
Column 3 gives the trial step size ($\delta \mathrm{DM}$), yielding the number of trials in Column 4 ($N_{\mathrm{DM}}$). 
Column 5 shows the downsampling factor ($d_s$), corresponding to the effective time resolution in Column 6 ($\Delta t_{\mathrm{eff}}$). 
Column 7 lists the number of frequency sub-bands ($n_{\mathrm{sub}}$) used in the sub-band de-dispersion algorithm \citep{Tremblay2015}.
\end{table}

Sensitivity was estimated using the standard radiometer equation:
\begin{equation}
\label{eq:sensitivity}
S_{\text{min}} = \frac{(S/N) \cdot T_{\text{sys}}}{G_{\rm inco} \sqrt{n_{\text{p}} t_{\text{int}} \Delta f}} \sqrt{\frac{\delta}{1 - \delta}},
\end{equation}
where $S/N$ is the minimum detectable signal-to-noise ratio, $T_{\text{sys}}$ is the system temperature, $G_{\rm inco}$ is the incoherent array gain, $n_{\text{p}}$ is the number of summed polarizations, $t_{\text{int}}$ is the integration time, $\Delta f$ is the observing bandwidth, and $\delta$ is the assumed pulse duty cycle. The calculation assumes $S/N = 4$, two polarizations, and a duty cycle of 3\%. The system temperature was taken as $T_{\text{sys}} = T_{\text{rec}} + T_{\text{sky}}$, with $T_{\text{rec}} = 23$ K \citep{Prabu2015} and $T_{\text{sky}}$ derived from the Full Embedded Element (FEE) model \citep{Sokolowski2017, Meyers2017}. Gain values were estimated following \citet{Oronsaye2015}, with an uncertainty of 10\%.

Applying this model yields the following sensitivity distribution: the survey has covered a total area of approximately 30,000 deg$^2$, achieving a best detection sensitivity of $\sim$8 mJy in the deepest regions, and averagely $\sim$40 mJy for the whole area (as shown in Figure \ref{fig:sensitivity}). Over 10,000 deg$^2$, the sensitivity reaches $S_{\rm min} < 0.1$ Jy, and detections in the far sidelobes indicate that sensitivity extends down to $\sim$1\% of the zenith gain $G_{\rm zenith} = 0.0274$ K Jy$^{-1}$. 

\section{Results}
\label{sec:results}

Our parallel search approaches have returned 80 confirmed detections, which are all known pulsars in the ATNF catalog \citep{Manchester2005}. Their sky distribution is shown in Figure~\ref{fig:sensitivity}, and information in Appendix Table ~\ref{appendix:80pulsars} with searched spin periods ($P_{0}$) ranging from 1.87 ms to 4.86 s and DMs between 2.64 and 180.16 pc cm$^{-3}$. Among these pulsars, 19 were found as harmonics, and six were detected via sidelobes of the incoherent beam—two (J2219+4754, J0332+5434) previously reported by \citet{Gong2020}, and four (J0736$-$6304, J1112$-$6926, J1141$-$6545, J1224$-$6407) newly identified—modestly increasing the effective sky coverage of the search. Our sample includes 4 MSPs, and 30 out of the 80 were detected for the first time at 185 MHz, as shown in the Appendix Table~\ref{appendix:80pulsars} and Figure~\ref{fig:30pulsars}.

\begin{figure*}
    \centering
    \includegraphics[width=14cm]{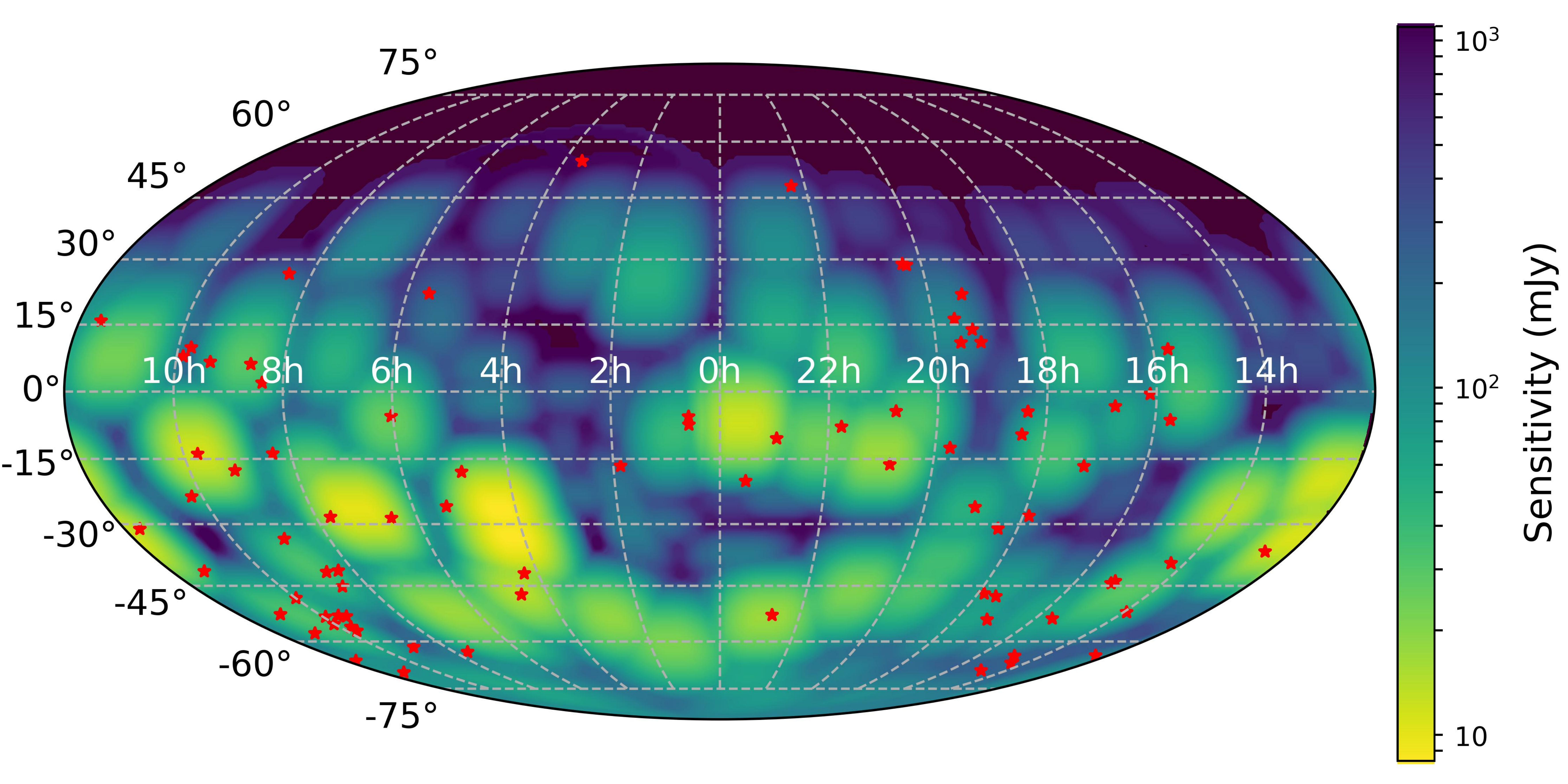}
    \caption{The sensitivity map includes sidelobes with beam power above 1\% of $G_{\text{zenith}}$. The detection threshold corresponds to $S/N_{\text{min}} = 4$, assuming a 3\% duty cycle. Our search methodology successfully identified 80 known pulsars marked in red stars.}
    \label{fig:sensitivity}
\end{figure*}

\subsection{Comparison with other low-frequency pulsar surveys}
\label{sec:comparison_surveys}

This search complements previous low-frequency pulsar efforts in both hemispheres. Within the MWA programme, \citet{Xue2017} conducted an initial incoherent census at 185 MHz. Our reprocessing of archival Phase I data extends this work through broader coverage and longer dwell times,increasing this previous census by 35 pulsars. By contrast, the SMART survey \citep{Bhat2023a, Bhat2023b, Lee2025} employs long coherent integrations with the Phase II compact configuration, achieving sensitivities of $\sim$2--3 mJy at 150 MHz and providing an extensive southern census that includes millisecond pulsars (MSPs). The GLEAM-X survey \citep{Mantovanini2025} adopts an imaging approach, delivering the first sub-400 MHz detections of more than 100 pulsars, and demonstrating the advantages of image-based searches for sources with extreme scattering or ultra-long periods. Together, these complementary strategies illustrate the trade-off between wide-area incoherent coverage, deep coherent searches, and image-based discovery.

Representative surveys in the northern hemisphere (e.g. AO327, GBNCC, LOTAAS, PUMPS) have extended the pulsar census above 100\,MHz with high sensitivity, while southern-hemisphere programmes (e.g. GHRSS) provide complementary coverage at comparable or lower frequencies. Collectively, these efforts span a wide range of observing frequencies, time and frequency resolutions, sky coverage, and sensitivities. Table~\ref{tab:survey_comparison} summarises their key parameters, placing the present MWA reprocessing in the broader survey landscape.

Taken together, these surveys highlight complementary strengths: incoherent MWA searches (this work; \citealt{Xue2017}) offer wide-area coverage with unparalleled fast computation, SMART provides deep sensitivity, and GLEAM-X imaging uncovers heavily scattered or ultra-long-period sources. Our reprocessing thus bridges the gap between wide-area census and deep targeted searches, reinforcing the importance of combining multiple approaches to achieve a comprehensive view of the low-frequency pulsar population.

\begin{table*}
\centering
\caption{Comparison of representative low-frequency pulsar surveys in the past decade.}
\label{tab:survey_comparison}
\begin{tabular}{lccccccc}
\hline
Survey & Telescope & Freq. band(MHz) & $t_{\rm res}$ ($\mu$s) &  $\Delta\nu$ (kHz)&  Sky coverage (deg$^2$) & $t_{\rm int}$ (s) & Sensitivity (mJy)\\
\hline
This work & MWA (Phase I) & 170--200 & 100 & 10 & 30,000 & 434--5075 & 8 \\
Xue+2017 & MWA (Phase I) & 170--200 & 100 & 10 & 17,000 & 384--5075 & -- \\
SMART & MWA (Phase II) & 140--170 & 100 & 10 & 31,000 & 4800 & 2--3 \\
GLEAM-X GP & MWA (imaging) & 72--231 & 500,000 & 10 & 3800 & 120 & 1--2 \\
GHRSS & GMRT & 306--338 & 30.72–61.44 & 15.63–31.25 & 2866 & 900, 1200 & 0.5\\
AO327 & Arecibo & 298--356 & 256/125/82 & 49/56/24 & $>18,000$ & 60 & $\lesssim$0.5\\
GBNCC & GBT & 300--400 & 82 & 24 & 35,100 & 120 & 0.7 \\
LOTAAS & LOFAR & 119--151 & 491.52 & 12.21 & 20,600 & 3600 & 1--2\\
PUMPS & LPA$^{a}$ & 110--112 & 12,500 & 78 & 20,100 & 210 & $>$0.1\\
\hline
\end{tabular}
\begin{flushleft}
\textbf{Note.} $^{a}$ Large Phased Array (LPA)
\end{flushleft}
\end{table*}

\subsection{The fluxes of the detected pulsars}
\label{sec:the fluxes of the detected pulsars}

At 185 MHz, we successfully measured the flux densities of 77 pulsars (Appendix Table~\ref{appendix:80pulsars}), with values ranging from 30 to 3600 mJy. For three pulsars—J0534+2200, J0835-4510, and J0034-0534—severe scattering smears the emission across nearly the entire rotation cycle. As no off-pulse region, could be identified to determine the baseline, reliable flux density estimates were unfeasible.

\begin{figure}
    \centering
    \includegraphics[width=8cm]{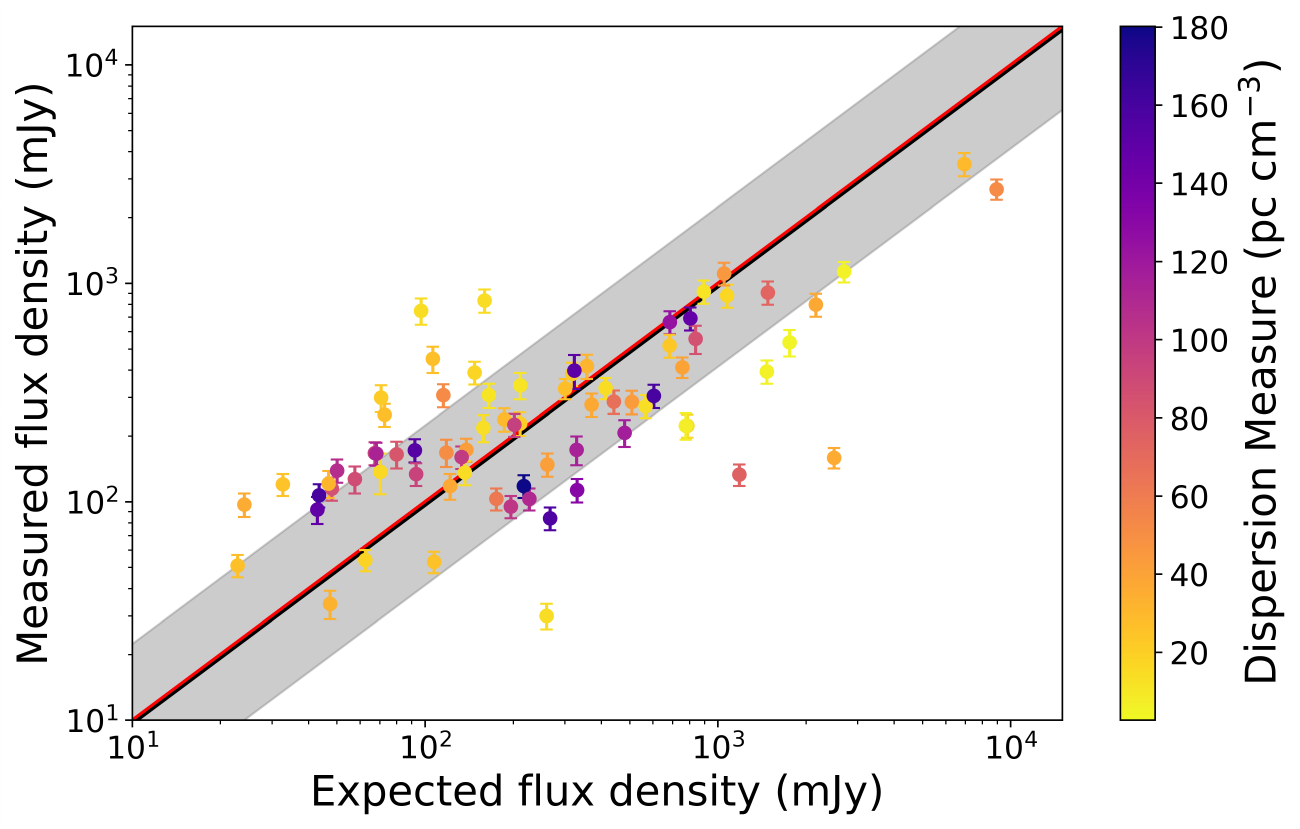}
    \caption{Comparison of measured 185 MHz flux densities ($S_{\text{185}}$) with expected values ($S_{\text{exp}}$) extrapolated from $\nu \geq 400$ MHz power-law spectra in the ATNF catalog. The red line represents unity ($S_{\text{185}} = S_{\text{exp}}$), and the grey band indicates the $1\sigma$ uncertainty range, centered on the best-fit ratio $S_{\text{185}}/S_{\text{exp}} = 0.96^{+0.11}_{-0.09}$} (black line).
    \label{fig:measured_flux_vs_expected_flux}
\end{figure}

By comparing the measured flux densities with the expected values extrapolated from 400 MHz catalog data \citep{Manchester2005}, assuming a spectral index of $\alpha = -1.6$, where no specific value was available\citep{Jankowski2018}, we obtained a median flux ratio of $0.96^{+0.11}_{-0.09}$. As shown in Figure \ref{fig:measured_flux_vs_expected_flux}, the measured flux densities are largely consistent with high-frequency extrapolations. Deviations from the 1:1 line are driven primarily by (1) spectral-model extrapolation uncertainties—several sources show curvature, broken spectra, or low-frequency turnover, so a single $\alpha$ inferred at $\gtrsim$400 MHz can over- or under-predict $S_{185}$ for individual objects \citep[e.g.][]{Jankowski2018,Sanidas2019,Bhat2023b,Bondonneau2020}—and (2) intrinsic flux variability on minute–hour timescales (e.g. mode changing, nulling, intermittency). In the ensemble, however, the agreement indicates that pronounced low-frequency turnover is uncommon above $\sim$180 MHz. This aligns with the 50--210 MHz range reported by \citet{Izvekova1981}.

The detection rate is satisfactory given the MWA-VCS's wide field of view and relatively short integration time. The search pipeline demonstrates robustness against radio frequency interference and sensitivity to sources in sidelobes, serving as an important complement to the direct folding method for known pulsars.

\subsection{Low-frequency spectral turnover}

For the detected pulsars, we combined our 185 MHz flux density measurements (Appendix Table~\ref{appendix:80pulsars}) with published values at other frequencies to construct broadband spectra. Spectral fitting was performed using the Python-based toolkit \texttt{pulsar\_spectra}\footnote{\url{https://github.com/NickSwainston/pulsar_spectra}} \citep{Swainston2022}, which implements Bayesian model selection across five models: (1) single power law; (2) broken power law; (3) power law with low-frequency turnover; (4) power law with high-frequency cutoff; and (5) double broken power law.

Model selection followed the Akaike Information Criterion (AIC), and 68\% confidence intervals were estimated via Markov Chain Monte Carlo (MCMC) sampling, following the methodology of \citet{Jankowski2018}. Among the five spectral models tested (see \citealt{Swainston2022} for definitions), the turnover model was preferred for 47 pulsars  (Appendix Table~\ref{appendix:all_models_final_results}), indicating robust evidence for low-frequency spectral turnovers in these sources. Since only a single flux density measurement at 185 MHz is contributed for each source, the additional point does not materially alter the AIC-based model selection, but it improves the robustness of the fitted parameters especially below 200 MHz.

\begin{figure*}
    \centering
    \includegraphics[width=\textwidth]{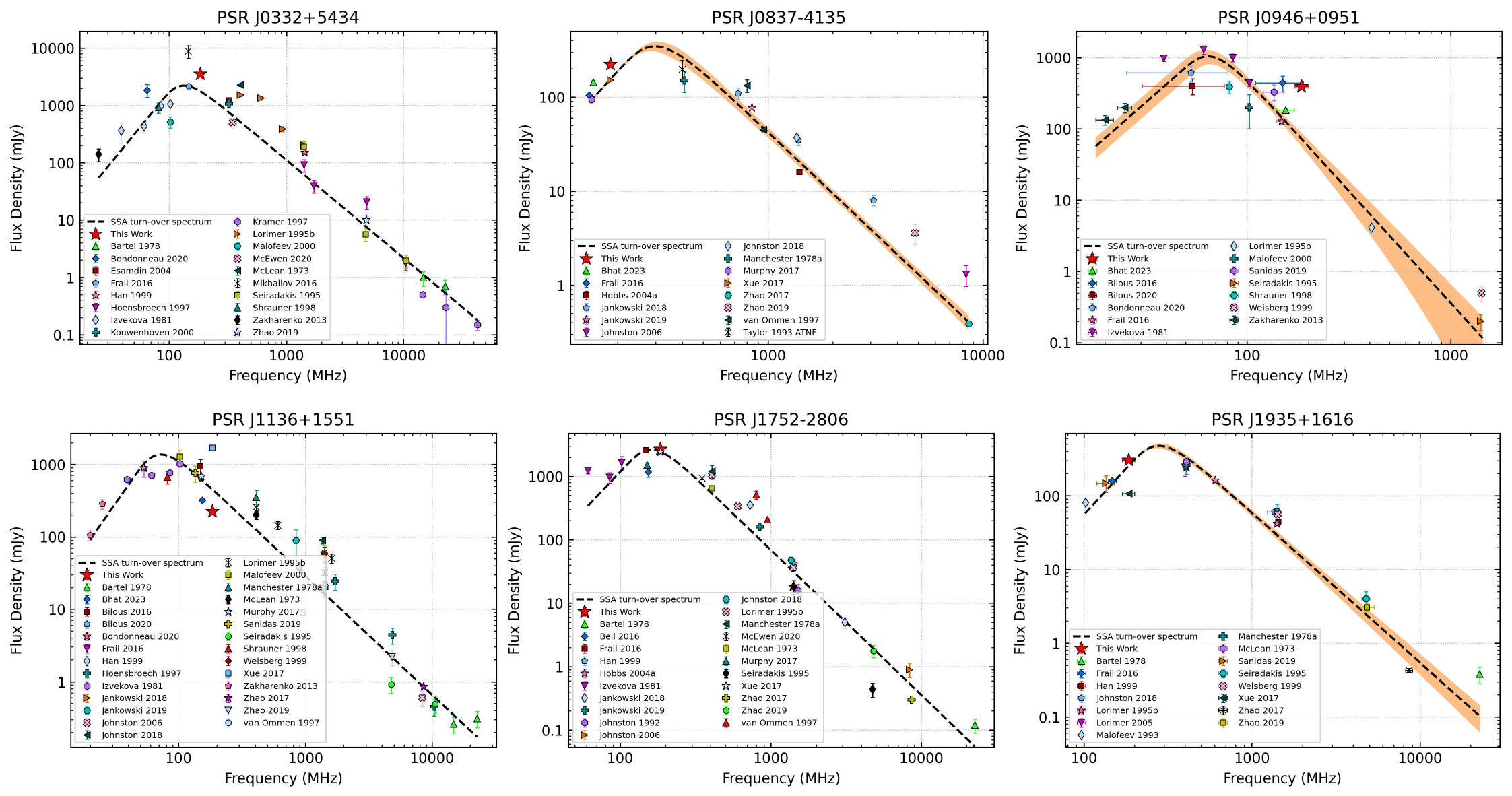}
    \caption{SSA model fits for six pulsars, showing the best-fit results based on archival and our measured flux densities}.
    \label{fig:ssa}
\end{figure*}

For these turnover candidates, we applied the synchrotron self-absorption (SSA) model approximation from \citet{Sieber1973}, conducting a systematic parameter space exploration. Six sources were found to be consistent with SSA within uncertainties (Figure~\ref{fig:ssa}).

\begin{figure}
    \centering
    \includegraphics[width=8cm]{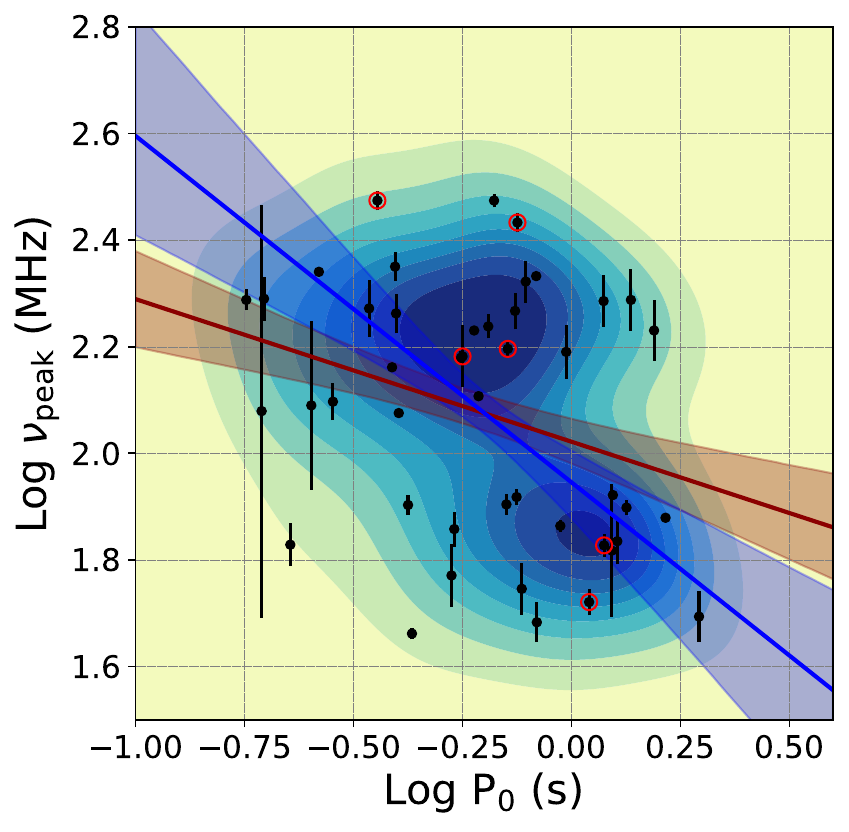}
    \caption{The shaded regions show 2D kernel density estimates (KDE). The red circles mark the sources with the best-fit SSA model. The red and blue lines indicate the OLS and the ORD fittings, respectively, with their $1\sigma$ uncertainties from bootstrapping.}
    \label{fig:p-vpeak}
\end{figure}

A possible correlation between turnover frequency $\nu_{\rm peak}$ and spin period $P_0$ (Figure~\ref{fig:p-vpeak}) was investigated using two regression methods: ordinary least-square \citep[OLS;][]{montgomery2012introduction}, which does not explicitly account for measurement uncertainties in both variables, and orthogonal distance regression \citep[ODR;][]{boggs1990odr}, which incorporates full error propagation. To avoid potential bias from MSPs, whose emission properties and spectral behaviour differ markedly from the general pulsar population \citep[e.g.][]{Kuniyoshi2015, Jankowski2018}, we excluded PSRs J0437–4715 and J2145–0750 from the fit. The OLS fit then yields a slope of $-0.268 \pm 0.127$ with $R^2 = 0.093$, while ODR gives $-0.657 \pm 1.264$ with $R^2 = -0.104$.

We note that for some sources the turnover fits yield large uncertainties on $\nu_{\rm peak}$. This reflects the limited flux sampling at low frequencies, where the spectral downturn is poorly constrained. In these cases, although the AIC still favours a turnover model over a simple power law, the resulting parameter posteriors are broad, leading to correspondingly large error bars. The apparent anti-correlation between $\nu_{\rm peak}$ and $P_0$ is therefore not statistically significant, and a substantially larger sample of long-period pulsars with well-constrained low-frequency flux densities will be required to robustly test this trend.

\subsection{Low-frequency scatter broadening}

At frequencies below 300 MHz, interstellar scattering causes significant temporal broadening of pulsar signals, leading to a decrease in sensitivity to short-period sources, especially MSPs situated in high-DM regions \citep{Lohmer2001,Lohmer2004,Lewandowski2013}. While dispersive smearing ($\Delta t_{\rm DM} \propto \nu^{-3}$) can be corrected via coherent or incoherent de-dispersion, pulse broadening due to multipath scattering ($\tau \propto \nu^{\alpha_{\rm sc}}$) is irreversible and often dominates at low observing frequencies.

Of the 80 pulsars detected in our census, 56 have published multi-frequency scattering measurements in the compilation of \citet{He2024}. For these sources, we estimated the expected scattering timescales $\tau_{\rm exp}$ at 185 MHz by extrapolating their reported spectral indices $\alpha_{\rm sc}$. Among these, 20 pulsars show $\tau_{\rm exp}$ exceeding the measured pulse width at 10\% of the peak ($W_{10}$), with approximately half satisfying $\tau_{\rm exp}/W_{10} > 0.3$, suggesting that scatter broadening significantly affects detectability in this search.

\begin{figure*}
    \centering
    \includegraphics[width=\textwidth]{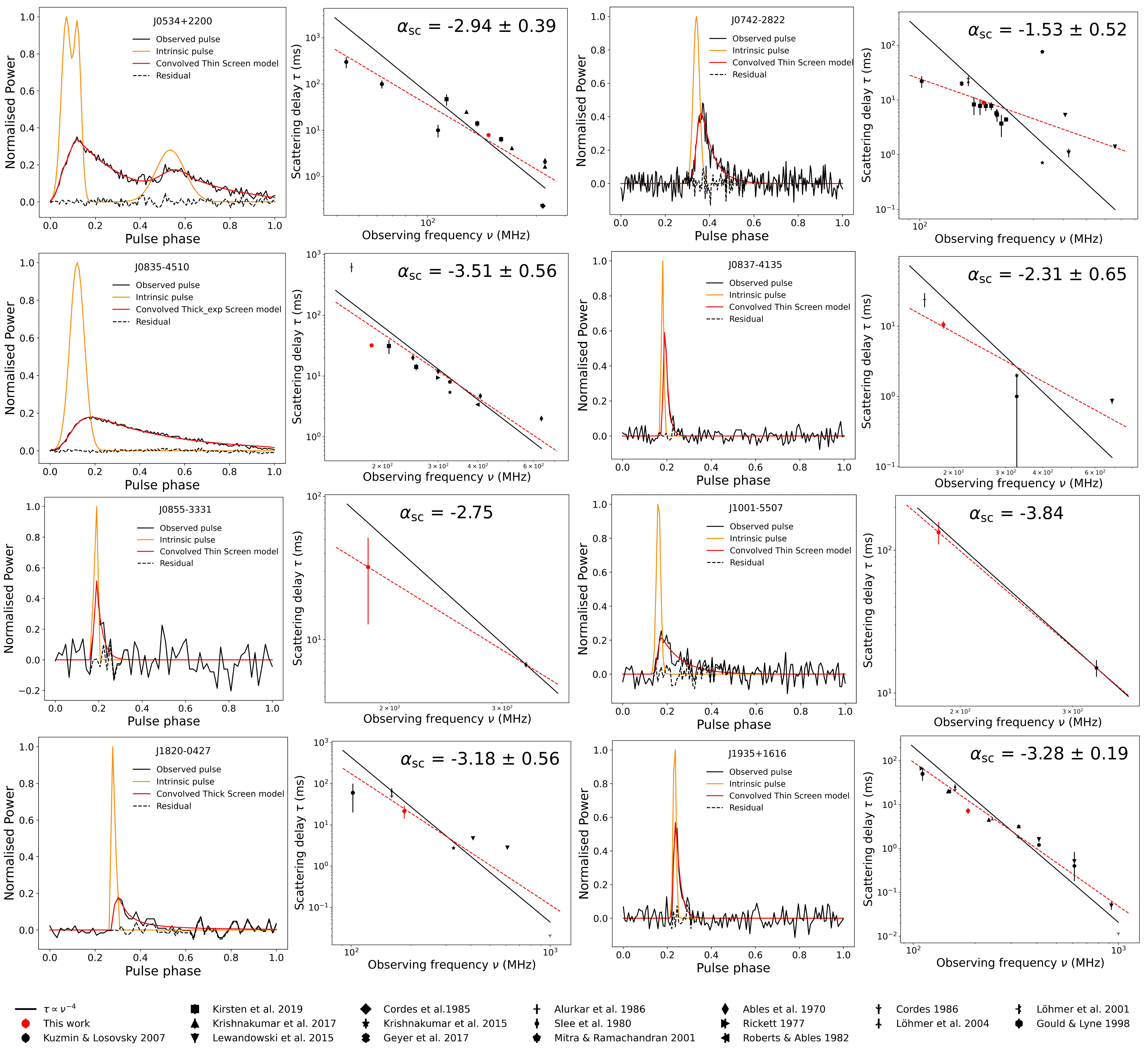}
    \caption{Forward-convolution fits for eight pulsars with resolvable scattering tails. Black curves show the observed profiles; red curves are the best-fit models obtained by convolving the assumed intrinsic profile (yellow) with thin- or thick-screen pulse broadening functions. Black dashed lines indicate residuals. Right panels present frequency-dependent scattering times: black points are literature values ($\nu < 1000$ MHz), red points are our 185-MHz measurements. The red dashed line shows the best-fit power-law relation $\tau \propto \nu^{\alpha_{\rm sc}}$, while the black solid line marks $\alpha_{\rm sc}=-4$ for reference.}
    \label{fig:scatter_pulsar}
\end{figure*}

For eight pulsars with resolvable scattering tails (see Table~\ref{tab:pbf_fits} and Figure~\ref{fig:scatter_pulsar}), we adopted a forward-convolution approach, in which the observed profile is expressed as the intrinsic pulse convolved with the ISM transfer function, dispersion smear, and instrumental response \citep[e.g.][]{Williamson1972, Krishnakumar2015}. For practical implementation these effects were combined into a single pulse broadening function (PBF),
\begin{equation}
I_{\rm obs}(t) = I_{\rm int}(t) \ast PBF(t),
\end{equation}
where $I_{\rm int}(t)$ denotes the intrinsic pulse shape. This method avoids the instabilities of deconvolution \citep[e.g.][]{Bhat2003}, with intrinsic profiles constrained iteratively from the data under varying levels of broadening. In a few complex cases (e.g. Vela), higher-frequency profiles provided loose guidance, but final parameters were obtained solely from fits at the observed frequency.

We considered two canonical forms of PBFs arising from single-screen scattering models \citep{Williamson1972, Williamson1973}:
\begin{equation}
PBF_1(t) = e^{-t/\tau} U(t), \quad
PBF_2(t) = \sqrt{\frac{\pi \tau}{4 t^3}} \exp\left(-\frac{\pi^2 \tau}{16 t}\right),
\end{equation}
where $t$ denotes the time lag relative to the onset of the scattered pulse, $\tau$ is the scattering timescale, and $U(t)$ is the unit step function. These represent scattering by a thin screen and a thick screen, respectively. Model fitting was performed by minimising residuals between the observed and convolved profiles at the survey frequency.

Six pulsars—J0742-2822, J0837-4135, J0855-3331, J1001-5507, J1935+1616, and J0534+2200—are adequately modelled by $PBF_1$. For these, the fitted spectral indices range from -3.84 to -1.53 (Table \ref{tab:pbf_fits}), satisfying with $\alpha_{\rm sc} > -4.0$, which are shallower than the canonical Kolmogorov turbulence expectation of $\alpha_{\rm sc} = -4.4$ \citep{Rickett1977, Romani1986}. This is consistent with previous findings that deviations from Kolmogorov scaling are common in regions with complex or non-uniform scattering media \citep[e.g.,][]{Lohmer2004, Geyer2017}.

The Crab pulsar J0534+2200 was likely modulated by wind-driven turbulence in the Crab Nebula \citep{Rickett1977}. For PSRs J0855–3331 and J1001–5507, $\alpha_{\rm sc}$ values of $-2.75$ and $-3.84$ were obtained, though these are based on only two frequency points and should be treated as indicative estimates pending further measurements. These may reflect either limitations in the frequency coverage of available data or environmental factors such as localised turbulence near H~\textsc{i} boundaries \citep{Koribalski1995}. PSR J1935+1616, located in the Galactic disk in agreement with previously reported anomalous scattering trends \citep{Lohmer2004}.

For PSR J1820-0427, the thin-screen model was inadequate. The thick-screen formulation provided a significantly better fit consistent with the CLEAN-based deconvolution result of \citet{Janagal2023}, who reported $\alpha_{\rm sc} = -3.50 \pm 0.30$.

The Vela pulsar presents scatter broadening inconsistent with either canonical model. We thus introduced a hybrid scattering model:
\begin{align}
w(t) &= 0.5 \left[1 + \tanh\left(\frac{0.09}{\tau} \cdot 30 (t - t_{\rm offset})\right)\right], \\
PBF_3(t) &= PBF_1(t) + w(t) \cdot (PBF_2(t) - PBF_1(t)),
\end{align}
which yielded $\alpha_{\rm sc} = -3.51 \pm 0.56$. This result aligns with \citet{Kirsten2019}, who reported that at 256 MHz the thick-screen model better fits the data, while at 210 MHz the thin-screen model was more appropriate.

\begin{table}
\centering
\caption{Best-fitting PBFs and scattering spectral indices for pulsars with resolvable scattering tails.}
\label{tab:pbf_fits}
\begin{tabular}{lcc}
\hline
Pulsar & PBF model & $\alpha_{\rm sc}$ \\
\hline
J0534+2200  & $PBF_1$   & $-2.94 \pm 0.39$ \\
J0742-2822 & $PBF_1$   & $-1.53 \pm 0.52$ \\
J0835-4510 & $PBF_3$   & $-3.51 \pm 0.56$ \\
J0837-4135 & $PBF_1$   & $-2.31 \pm 0.65$ \\
J0855--3331 & $PBF_1$   & $-2.75$ \\
J1001--5507 & $PBF_1$   & $-3.84$ \\
J1820--0427 & $PBF_2$   & $-3.18 \pm 0.56$ \\
J1935+1616  & $PBF_1$   & $-3.28 \pm 0.19$ \\
\hline
\end{tabular}
\end{table}

\subsection{Pulse width–period relations}

The pulse width $W$ provides direct observational constraints on the geometry and radiation mechanisms of pulsars \citep{Gil1984}. Under the classical cone-shaped beam model, assuming a magnetic inclination angle $\alpha = 90^\circ$ and an impact angle $\beta = 0^\circ$ (hence the dispersion beyond the intrinsic constraints on the following relations are mainly caused by the random distribution of both angles), the observed pulse width $W$ is related to the emission beam radius $\rho$ by $W = 2\rho$. If the emission originates from a constant height $r_{\rm em}$, this leads to $W \propto P^{-0.5}$ \citep{Rankin1990, Rankin1993}. Accounting for the $P$- and $\dot{P}$-dependent emission height \citep{Kijak1998, Kijak2003}, and incorporating the empirical $\dot{P} \propto P^{2-3}$ scaling \citep{Lyne1985, Hobbs2005, Faucher2006}, one obtains $W \propto P^{-0.26\pm0.02}$ \citep{Lorimer2004}.

Given the increased influence of $r_{\rm em}$ at low frequencies, our data are well suited for investigating such dependencies. We measured $W_{10}$ using multi-Gaussian fitting, incorporating uncertainties from the sampling time ($t_{\text{bin}}$), scattering timescale ($\tau_{\rm sc}$), and intra-channel dispersion delay ($\Delta t_{\text{chan}}$). Pulsars with strong scattering (J0534+2200, J0835-4510) or multi-component profiles (J0820-4114, J0959-4809, J1057-5226, J2145-0750) were excluded to minimize systematic biases.

We performed regression analysis on $W_{10}$ and $P_0$ using both OLS and ODR. The resulting power-law indices were $\mu_{\text{OLS}} = -0.25 \pm 0.04$ and $\mu_{\text{ODR}} = -0.28 \pm 0.07$, with consistent determination coefficients ($R^2 \sim 0.32$). These results agree with previous studies at various frequencies, including 1400 MHz \citep[$-0.28 \pm 0.03$,][]{Johnston2019}, 1284 MHz \citep[$-0.29 \pm 0.03$,][]{Posselt2021}, 350 MHz \citep[$-0.27 \pm 0.001$,][]{McEwen2020}, and 150 MHz \citep[$-0.3 \pm 0.4$,][]{Pilia2016}.

We further investigated the dependence of $W_{10}$ on $\dot{P}$, spin-down energy ($\dot{E}$), and surface magnetic field strength ($B_{\text{surf}}$), with details provided in Table~\ref{tbl:w_p}. The regression results indicate that $P_0$ has the strongest correlation with $W_{10}$, while $\dot{P}$ shows the weakest. The $\dot{E}$--$W_{10}$ regression is positive ($\mu_{\text{OLS}} = 0.12 \pm 0.03$, $R^2 = 0.21$), indicating a trend of broader pulse widths for pulsars with higher spin-down energy.

\begin{table}
\centering
\caption{OLS and ODR Regression Results of $W_{10}$ with Parameters}
\label{tbl:w_p}
\begin{tabular}{lcccc}
\hline
Parameter & \multicolumn{2}{c}{OLS} & \multicolumn{2}{c}{ODR} \\
\cline{2-3} \cline{4-5}
          & $\mu$ & $R^2$ & $\mu$ & $R^2$ \\
\hline
$P_0$                & $-0.25\pm0.04$ & 0.324 & $-0.28\pm0.07$ & 0.318 \\
$\dot{P}$            & $-0.08\pm0.02$ & 0.169 & $-0.08\pm0.03$ & 0.169 \\
$\dot{E}$            & $0.12\pm0.03$  & 0.214 & $0.13\pm0.03$  & 0.213 \\
$B_{\text{surf}}$    & $-0.13\pm0.03$ & 0.227 & $-0.14\pm0.05$ & 0.226 \\
\hline
\end{tabular}
\end{table}

\begin{figure*}
    \centering
    \includegraphics[width=\textwidth]{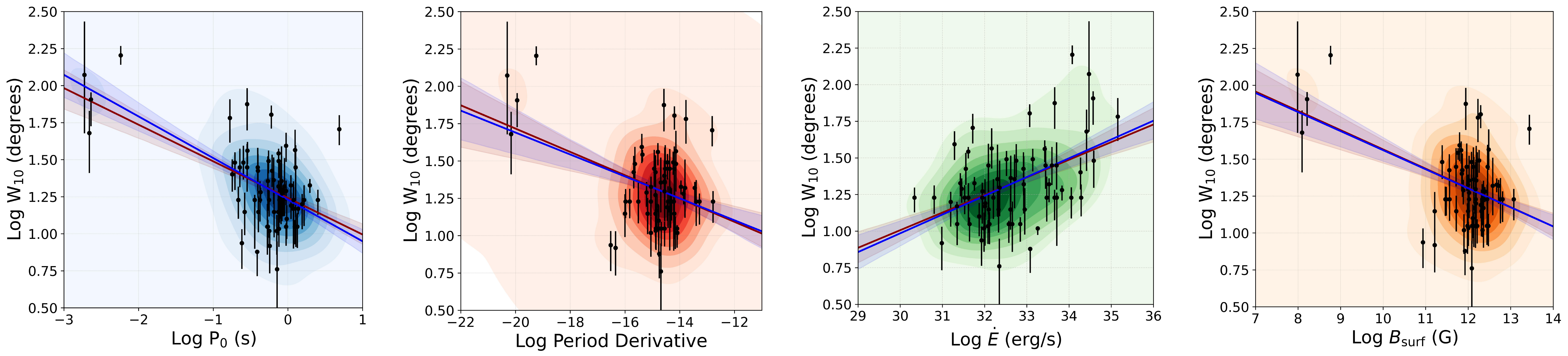}
    \caption{Log-log plots of $W_{10}$ relations with physical parameters. The sample values are in black dots, with the shaded contours showing 2D KDE. The red and blue lines indicate the OLS and the ORD fittings, respectively, with their $1\sigma$ uncertainties from bootstrapping.}
    \label{fig:pulsar_width}
\end{figure*}

\section{Discussion}
\label{sec:discussion}

\subsection{The Undetected Pulsars}
\label{sec:undetected pulsar}

\begin{figure*}
    \centering
    \includegraphics[width=15cm]{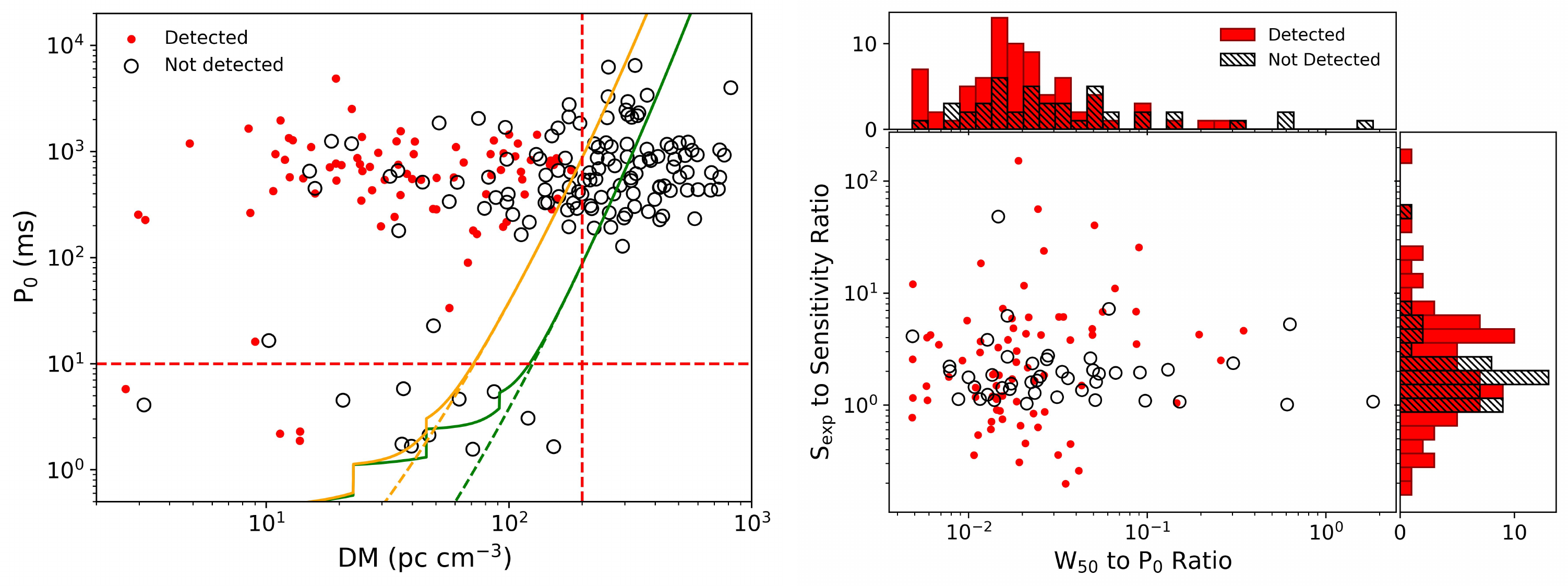}
    \caption{Left: $P_0$-DM distribution of detected (filled circles) and undetected (open circles) pulsars. green and orange lines show thresholds from one and ten times of pulse scatter broadening \citep{Bhat2004}, respectively with (solid) or without (dashed) DM smearing. Right: $W_{50}/P_0$ versus flux ratio ($S_{\text{exp}}/S_{\text{min}}$) of pulsars above the solid orange line in the left panel.}
    \label{fig:sensitivity_dm_p}
\end{figure*}

To quantify the search completeness, a systematic comparison was made between the more accurately predicted 185 MHz flux densities of known pulsars (derived via the \texttt{pulsar\_spectra}) and the local sensitivity limits. Figure \ref{fig:sensitivity_dm_p} shows that 105 pulsars with predicted fluxes above the detection thresholds were not detected, indicating two main selection biases. High-DM pulsars (DM $\gtrsim$150 pc cm$^{-3}$, to the right of the orange solid line in the left panel of Figure \ref{fig:sensitivity_dm_p}) experience scattering-induced pulse broadening, which follows $\tau_{\text{sc}} \propto \text{DM}^{2}\nu^{-4}$ \citep{Bhat2004}. For such sources, the effective pulse width $W_{\text{eff}} = \sqrt{W_{\text{int}}^2 + \tau_{\text{sc}}^2}$ exceeds $0.1P_0$, leading to a reduction in FFT-based detectability due to the $S/N \propto W_{\text{eff}}^{-1/2}$ scaling  relationship. This accounts for the non-detection of approximately 70\% of the theoretically detectable population with DM $>$ 150 pc cm$^{-3}$, consistent with LOFAR low-frequency constraints \citep{Bilous2020}. 

Meanwhile, the undetected low-DM pulsars (to the left of the orange line) are clustered near the sensitivity floor ($S_{\text{185}}/S_{\text{min}} \sim$1-3, shown in the right panel of Figure \ref{fig:sensitivity_dm_p}). The non-detection of these pulsars may be attributed to several factors. Firstly, at low frequencies, most low-DM pulsars are expected to lie in the strong-scintillation regime, with scintillation indices $m \sim 1$ \citep[e.g.][]{Rickett1990}, corresponding to flux modulations of order unity. For observations with finite bandwidth and integration time the effective $m$ is reduced, but variations of tens of per cent remain plausible and can readily shift marginal sources below or above the detection threshold \citep{Wang2001,Wang2005}. Secondly, there may be unexpected spectral turnovers below 200 MHz that were not considered in high-frequency flux extrapolations \citep{Jankowski2018}. Thirdly, their systematically broader $W_{50}/P_0$ ratios (with a median value of 0.028 compared to 0.018 for detected pulsars) make their signals less detectable in FFT-based searches. 

In addition to these factors, the non-detections could also be influenced by the pulsars' low-frequency emission characteristics. Some pulsars might have intrinsically weak or complex low-frequency emission that is not well-represented by the extrapolated models. Furthermore, the observational strategy could be optimized to enhance detection rates. This includes refining the search algorithms to better handle the diverse pulse profiles and scattering effects at low frequencies, as well as increasing the integration time or observing bandwidth to improve sensitivity. A more comprehensive analysis incorporating these aspects would provide a fuller explanation for the non-detections.

\subsection{Spectral turnovers in the low-frequency regime}
\label{sec:discussion_spectra}

The identification of 47 pulsars with low-frequency spectral turnovers reinforces the view that many pulsar spectra deviate from simple power laws at metre wavelengths \citep{Izvekova1981, Jankowski2018}. These results expand the sample of pulsars exhibiting low-frequency spectral turnovers in the $\sim$100--400~MHz range.

Only six of the 47 sources exhibit spectra consistent with the SSA model of \citet{Sieber1973}, suggesting that SSA alone cannot explain most turnover behavior. While SSA is physically motivated in compact, magnetized plasmas, its applicability to the extended pulsar magnetosphere remains uncertain. Alternative mechanisms like free-free absorption (FFA) by surrounding media--including supernova remnants, H~\textsc{ii} regions, or dense filaments in the ISM--are likely to play a role \citep{Kijak2021}. 

However, the relative contributions of these mechanisms and their specific conditions of applicability warrant deeper exploration. SSA is typically effective in compact, magnetized plasmas, but may be less dominant in the extended magnetospheric regions where pulsar emission originates. FFA, on the other hand, may be more significant in environments such as supernova remnants or H~\textsc{ii} regions surrounding some pulsars. The diversity of observed spectral turnovers suggests that multiple absorption processes may operate simultaneously, with their relative importance varying among different pulsars.

The tentative anti-correlation between $\nu_{\text{peak}}$ and $P_0$, as shown in Figure~\ref{fig:p-vpeak}, suggests that longer-period pulsars might exhibit spectral turnovers at lower frequencies. This aligns with theoretical expectations that pulsars with lower spin-down energy could emit more efficiently from regions with plasma densities conducive to absorption. The negative slopes derived from both OLS ($-0.26\pm0.12$) and ODR ($-0.52\pm0.24$) analyses support this trend, though the statistical significance remains weak ($\lesssim 2\sigma$) with slope uncertainties exceeding 45\% and low $R^2$ values ($\sim0.1$).

These limitations likely reflect a combination of factors including modest sample size, unmodelled intrinsic variations (e.g., viewing geometry effects, spectral variability), and observational selection biases. To better understand the roles of different absorption mechanisms, future studies should incorporate more comprehensive multi-frequency observational data across a wider spectral range. This should be coupled with improvements to theoretical models of pulsar emission and propagation to better account for magnetospheric and interstellar medium effects. Additionally, expanding samples with well-characterized spectra and independently constrained emission geometries will be crucial for establishing robust correlations between pulsar properties and their spectral turnover characteristics.

Multi-frequency polarimetric observations will be particularly valuable for several key investigations. First, they will help disentangle SSA from competing absorption processes like FFA by enabling detailed analysis of frequency-dependent polarization signatures. Second, such observations can probe magnetospheric influences on spectral turnover behavior through correlated studies of rotation measure variations and profile evolution across frequencies. Third, they will provide critical constraints on the spatial distribution of absorbing media by mapping Faraday rotation and scattering screens along different lines of sight to pulsars. The combination of these approaches will yield a more complete picture of low-frequency pulsar emission physics and the intervening absorption processes that shape observed spectra.

Such advances will help establish a more complete picture of low-frequency pulsar emission and the intervening absorption processes that shape their observed spectra.

\subsection{Scattering properties and propagation path complexity}
\label{sec:discussion_scattering}

Scatter broadening remains a significant impediment in low-frequency pulsar surveys, particularly affecting fast-spinning and distant sources. The detection threshold characterized by $\tau_{\rm exp}/W_{10} > 0.3$ encompasses a considerable fraction of the sample, indicating that even moderate scattering substantially reduces pulsar detectability at 185~MHz.

Most fitted spectral indices $\alpha_{\rm sc}$ were flatter than the expected Kolmogorov value, indicating that the underlying scattering medium often deviates from homogeneous turbulence. Such deviations, especially in regions like the Crab Nebula or the Gum Nebula, likely reflect complex plasma structures, inhomogeneous density distributions, or varying magnetic field geometries \citep{Rickett1977, Geyer2017, Lohmer2004}.

The need for a hybrid model in the case of Vela suggests that standard thin/thick-screen models are insufficient in multi-phase environments, where both large-scale gradients and dense localized clumps may coexist \citep{Cantat-Gaudin2019, Kirsten2019}. Our hybrid PBF bridges these scenarios by interpolating between limiting cases, offering a more flexible empirical tool for forward-modeling scattering features.

These results reinforce the diagnostic power of pulse profile modeling at low frequencies. Forward convolution methods, unlike deconvolution techniques, remain robust at low S/N and permit physical interpretation of scattering structures along different lines of sight. With improved data quality and future polarimetric follow-up, more accurate mapping of scattering geometries will be possible—critical for refining Galactic electron density models and ISM turbulence frameworks.

\subsection{Pulse width--period relation at low frequencies}
\label{sec:discussion_width}

These findings confirm the established inverse relationship between pulse width and spin period. This is consistent with emission originating at higher altitudes in shorter-period pulsars, where the open field line region subtends a larger angle. The derived power-law index ($\mu \approx -0.25$ to $-0.28$) agrees with previous studies across a wide frequency range, confirming that this trend holds in the low-frequency regime.

The weaker dependence of $W_{10}$ on $\dot{P}$ likely reflects both the limited dynamic range of this parameter within the sample and the subordinate influence of magnetic braking on beam morphology. The modest positive correlation between $W_{10}$ and $\dot{E}$ may be interpreted in two ways: either high-$\dot{E}$ pulsars exhibit magnetospheric expansion and beam broadening due to greater energy availability, or the correlation reflects an evolutionary sequence in which younger, high-$\dot{E}$ pulsars possess wider beams due to their less confined magnetospheres and larger light cylinder radii ($R_{\text{LC}} \propto P$). A corresponding decrease in $B_{\text{surf}}$ along with increasing $W_{10}$ could be consistent with such an evolutionary scenario \citep{Li2023, Li2024}.

Despite the overall robustness of the $W_{10}$–$P_0$ relation, the scatter in the regression highlights the influence of geometric and physical factors not included in this simple model, including beam shape variations, inclination angle, and frequency-dependent emission height. More precise estimates of $r_{\rm em}$, possibly through polarimetric studies or emission altitude mapping, will help disentangle these effects in future work.

\section{Conclusions and Prospectives}
\label{sec:conclusions}
Our MWA VCS search represents a significant advance in low-frequency pulsar astronomy, covering 30,000 deg$^2$ of the southern sky at 185 MHz with the best sensitivity of 8 mJy. The catalog of 80 pulsars, including 30 with first-time measurements at this frequency \citep{Wei2023}, provides an unprecedented resource for understanding pulsar emission mechanisms and interstellar medium properties. These results establish essential benchmarks for the upcoming SKA era while also revealing important avenues for methodological refinement.

\subsection{Key Findings}
The spectral analysis revealed that 47 pulsars show significant low-frequency turnovers between 100-400 MHz. While only six fit standard SSA models, most cases suggest alternative mechanisms like FFA dominate. We observed a tentative anti-relation between turnover frequency and spin period, though larger samples are needed for confirmation due to selection effects.

Scattering analysis demonstrated that interstellar effects strongly shape low-frequency emission, with spectral indices generally shallower than Kolmogorov turbulence predictions. This indicates complex multi-phase interstellar medium structure. Special cases like the Vela and Crab pulsars required hybrid scattering models to accurately describe their profile broadening.

The single-frequency nature of our 185 MHz observations presents inherent limitations for spectral analysis. Without  simultaneous multi-frequency data, we cannot fully distinguish between intrinsic pulsar emission properties, instrumental  systematics, and interstellar propagation effects. This particularly impacts our understanding of spectral turnover mechanisms and frequency-dependent scattering. The scattering models we employed, while computationally efficient, may oversimplify the actual interstellar medium structure, especially in turbulent regions.

\subsection{Methodological Reflections}

The constraints identified in Sections \ref{sec:undetected pulsar}---\ref{sec:discussion_spectra} collectively highlight three fundamental trade-offs in low-frequency pulsar surveys: (1) spectral resolution versus observational efficiency at 185 MHz, (2) sensitivity thresholds (8 mJy) versus population completeness, and (3) computational tractability versus ISM complexity in scattering models. Rather than limiting the study's validity, these trade-offs explicitly define the parameter space for future SKA-Low observations.

Specifically, the single-band observations that constrained our spectral analysis now serve as a baseline for designing multi-frequency campaigns with LOFAR (150 MHz) and uGMRT (300 MHz). Similarly, the hybrid scattering models developed for extreme cases like the Vela pulsar demonstrate a scalable framework for SKA's wider field of view observations. This transformative potential outweighs the initial constraints, as evidenced by our catalog's utility as calibration sources for next-generation surveys

\subsection{Future Prospects}
Looking ahead, three strategic directions emerge from this work. First, coordinated observations with low-frequency facilities such as LOFAR (150 MHz), the MWA (185 MHz), and uGMRT (300 MHz) can refine measurements of scattering and spectral turnovers. More stringent constraints, however, require joint campaigns with higher-frequency telescopes (e.g. Parkes, FAST, Effelsberg), providing coverage from $\sim$100 MHz to several GHz. Such broad-band studies are critical for disentangling intrinsic emission properties from propagation effects by the ISM. Second, deeper integrations could specifically target the high-DM, non-detected pulsars in our sample to directly quantify the scattering and constrain the extreme ISM properties along those lines of sight, thereby revealing populations of weak low-frequency emitters below our current sensitivity threshold. Third, systematic polarization studies would enable more robust constraints on emission geometry and its relationship to scattering phenomena.

For next-generation facilities like SKA-Low, this study provides crucial preparatory foundations. The data products serve as valuable calibration sources, while the analysis methods offer templates for large-scale survey strategies. Particularly significant is our development of hybrid scattering models, which demonstrate an approach that could be adapted to handle the complex interstellar environments expected in SKA observations.

This work establishes a comprehensive framework for advancing low-frequency pulsar astronomy. By combining our observational results with methodological innovations and clear pathways for future research, we have laid substantial groundwork for the next era of discovery. The coming years promise significant breakthroughs in understanding pulsar emission physics, interstellar medium structure, and extreme astrophysical environments as these research directions are pursued. 


\section*{Acknowledgements}

We acknowledge the support of the China SKA Regional Centre Prototype for the provided computational resources. We also thank Nick Swainston, Mengyao Xue, and Bradley Meyers from the MWA team for their technical guidance. This work was supported by the National SKA Program of China (No. 2020SKA0120201), the Major Science and Technology Program of Xinjiang Uygur Autonomous Region (No. 2022A03013-1), National Natural Science Foundation of China (No. 12041304, No. 12288102 and No. 12373114, 12003009).

\section*{Data Availability}

The raw MWA-VCS data used in this work are available through the MWA All-Sky Virtual Observatory (ASVO) archive at \url{https://asvo.mwatelescope.org}.  
The pulsar profiles, flux measurements, and derived scattering parameters will be shared upon reasonable request to the corresponding author.  
All data underlying this article are presented within the paper. Additional materials are not publicly available but can be provided upon reasonable request to the corresponding author. The custom code developed for the hybrid scattering model analysis is available from the corresponding author upon reasonable request.

\bibliographystyle{mnras}
\bibliography{example} 

\begin{thebibliography}{}
\makeatletter
\relax
\def\mn@urlcharsother{\let\do\@makeother \do\$\do\&\do\#\do\^\do\_\do\%\do\~}
\def\mn@doi{\begingroup\mn@urlcharsother \@ifnextchar [ {\mn@doi@} {\mn@doi@[]}}
\def\mn@doi@[#1]#2{\def\@tempa{#1}\ifx\@tempa\@empty \href {http://dx.doi.org/#2} {doi:#2}\else \href {http://dx.doi.org/#2} {#1}\fi \endgroup}
\def\mn@eprint#1#2{\mn@eprint@#1:#2::\@nil}
\def\mn@eprint@arXiv#1{\href {http://arxiv.org/abs/#1} {{\tt arXiv:#1}}}
\def\mn@eprint@dblp#1{\href {http://dblp.uni-trier.de/rec/bibtex/#1.xml} {dblp:#1}}
\def\mn@eprint@#1:#2:#3:#4\@nil{\def\@tempa {#1}\def\@tempb {#2}\def\@tempc {#3}\ifx \@tempc \@empty \let \@tempc \@tempb \let \@tempb \@tempa \fi \ifx \@tempb \@empty \def\@tempb {arXiv}\fi \@ifundefined {mn@eprint@\@tempb}{\@tempb:\@tempc}{\expandafter \expandafter \csname mn@eprint@\@tempb\endcsname \expandafter{\@tempc}}}

\bibitem[\protect\citeauthoryear{{An}, {Wu}  \& {Hong}}{{An} et~al.}{2019}]{An2019}
{An} T.,  {Wu} X.-P.,   {Hong} X.,  2019, \mn@doi [Nat. Astron] {10.1038/s41550-019-0943-4}, \href {https://ui.adsabs.harvard.edu/abs/2019NatAs...3.1030A} {3, 1030}

\bibitem[\protect\citeauthoryear{{An}, {Wu}, {Lao}, {Guo}, {Xu}, {Lv}, {Zhang}  \& {Zhang}}{{An} et~al.}{2022}]{An2022}
{An} T.,  {Wu} X.,  {Lao} B.,  {Guo} S.,  {Xu} Z.,  {Lv} W.,  {Zhang} Y.,   {Zhang} Z.,  2022, \mn@doi [Sci. China Phys. Mech. Astron] {10.1007/s11433-022-1981-8}, \href {https://ui.adsabs.harvard.edu/abs/2022SCPMA..6529501A} {65, 129501}

\bibitem[\protect\citeauthoryear{{Bhat}, {Cordes}  \& {Chatterjee}}{{Bhat} et~al.}{2003}]{Bhat2003}
{Bhat} N.~D.~R.,  {Cordes} J.~M.,   {Chatterjee} S.,  2003, \mn@doi [\apj] {10.1086/345775}, \href {https://ui.adsabs.harvard.edu/abs/2003ApJ...584..782B} {584, 782}

\bibitem[\protect\citeauthoryear{{Bhat}, {Cordes}, {Camilo}, {Nice}  \& {Lorimer}}{{Bhat} et~al.}{2004}]{Bhat2004}
{Bhat} N.~D.~R.,  {Cordes} J.~M.,  {Camilo} F.,  {Nice} D.~J.,   {Lorimer} D.~R.,  2004, \mn@doi [ApJ] {10.1086/382680}, \href {https://ui.adsabs.harvard.edu/abs/2004ApJ...605..759B} {605, 759}

\bibitem[\protect\citeauthoryear{{Bhat} et~al.,}{{Bhat} et~al.}{2023a}]{Bhat2023b}
{Bhat} N.~D.~R.,  et~al., 2023a, \mn@doi [PASA] {10.1017/pasa.2023.18}, \href {https://ui.adsabs.harvard.edu/abs/2023PASA...40...20B} {40, e020}

\bibitem[\protect\citeauthoryear{{Bhat} et~al.,}{{Bhat} et~al.}{2023b}]{Bhat2023a}
{Bhat} N.~D.~R.,  et~al., 2023b, \mn@doi [PASA] {10.1017/pasa.2023.17}, \href {https://ui.adsabs.harvard.edu/abs/2023PASA...40...21B} {40, e021}

\bibitem[\protect\citeauthoryear{{Bhattacharyya} et~al.,}{{Bhattacharyya} et~al.}{2016}]{Bhattacharyya2016}
{Bhattacharyya} B.,  et~al., 2016, \mn@doi [\apj] {10.3847/0004-637X/817/2/130}, \href {https://ui.adsabs.harvard.edu/abs/2016ApJ...817..130B} {817, 130}

\bibitem[\protect\citeauthoryear{{Bhattacharyya} et~al.,}{{Bhattacharyya} et~al.}{2019}]{Bhattacharyya2019}
{Bhattacharyya} B.,  et~al., 2019, \mn@doi [\apj] {10.3847/1538-4357/ab2bf3}, \href {https://ui.adsabs.harvard.edu/abs/2019ApJ...881...59B} {881, 59}

\bibitem[\protect\citeauthoryear{{Bilous} et~al.,}{{Bilous} et~al.}{2020}]{Bilous2020}
{Bilous} A.~V.,  et~al., 2020, \mn@doi [\aap] {10.1051/0004-6361/201936627}, \href {https://ui.adsabs.harvard.edu/abs/2020A&A...635A..75B} {635, A75}

\bibitem[\protect\citeauthoryear{{Blandford} \& {Narayan}}{{Blandford} \& {Narayan}}{1985}]{Blandford1985}
{Blandford} R.,  {Narayan} R.,  1985, \mn@doi [\mnras] {10.1093/mnras/213.3.591}, \href {https://ui.adsabs.harvard.edu/abs/1985MNRAS.213..591B} {213, 591}

\bibitem[\protect\citeauthoryear{Boggs \& Rogers}{Boggs \& Rogers}{1990}]{boggs1990odr}
Boggs P.~T.,  Rogers J.~E.,  1990, Technical Report NISTIR 89-4197, Orthogonal Distance Regression.
National Institute of Standards and Technology, Gaithersburg, MD

\bibitem[\protect\citeauthoryear{{Bondonneau}, {Grie{\ss}meier}, {Theureau}, {Bilous}, {Kondratiev}, {Serylak}, {Keith}  \& {Lyne}}{{Bondonneau} et~al.}{2020}]{Bondonneau2020}
{Bondonneau} L.,  {Grie{\ss}meier} J.~M.,  {Theureau} G.,  {Bilous} A.~V.,  {Kondratiev} V.~I.,  {Serylak} M.,  {Keith} M.~J.,   {Lyne} A.~G.,  2020, \mn@doi [\aap] {10.1051/0004-6361/201936829}, \href {https://ui.adsabs.harvard.edu/abs/2020A&A...635A..76B} {635, A76}

\bibitem[\protect\citeauthoryear{{Bowman} et~al.,}{{Bowman} et~al.}{2013}]{Bowman2013}
{Bowman} J.~D.,  et~al., 2013, \mn@doi [PASA] {10.1017/pas.2013.009}, \href {https://ui.adsabs.harvard.edu/abs/2013PASA...30...31B} {30, e031}

\bibitem[\protect\citeauthoryear{{Cantat-Gaudin}, {Mapelli}, {Balaguer-N{\'u}{\~n}ez}, {Jordi}, {Sacco}  \& {Vallenari}}{{Cantat-Gaudin} et~al.}{2019}]{Cantat-Gaudin2019}
{Cantat-Gaudin} T.,  {Mapelli} M.,  {Balaguer-N{\'u}{\~n}ez} L.,  {Jordi} C.,  {Sacco} G.,   {Vallenari} A.,  2019, \mn@doi [\aap] {10.1051/0004-6361/201834003}, \href {https://ui.adsabs.harvard.edu/abs/2019A&A...621A.115C} {621, A115}

\bibitem[\protect\citeauthoryear{{Cordes} \& {Lazio}}{{Cordes} \& {Lazio}}{2002}]{Cordes2002}
{Cordes} J.~M.,  {Lazio} T.~J.~W.,  2002, \mn@doi [arXiv e-prints] {10.48550/arXiv.astro-ph/0207156}, \href {https://ui.adsabs.harvard.edu/abs/2002astro.ph..7156C} {pp astro--ph/0207156}

\bibitem[\protect\citeauthoryear{{Deneva}, {Stovall}, {McLaughlin}, {Bates}, {Freire}, {Martinez}, {Jenet}  \& {Bagchi}}{{Deneva} et~al.}{2013}]{Deneva2013}
{Deneva} J.~S.,  {Stovall} K.,  {McLaughlin} M.~A.,  {Bates} S.~D.,  {Freire} P.~C.~C.,  {Martinez} J.~G.,  {Jenet} F.,   {Bagchi} M.,  2013, \mn@doi [\apj] {10.1088/0004-637X/775/1/51}, \href {https://ui.adsabs.harvard.edu/abs/2013ApJ...775...51D} {775, 51}

\bibitem[\protect\citeauthoryear{{Deneva} et~al.,}{{Deneva} et~al.}{2016}]{Deneva2016}
{Deneva} J.~S.,  et~al., 2016, \mn@doi [\apj] {10.3847/0004-637X/821/1/10}, \href {https://ui.adsabs.harvard.edu/abs/2016ApJ...821...10D} {821, 10}

\bibitem[\protect\citeauthoryear{{Dewdney}, {Hall}, {Schilizzi}  \& {Lazio}}{{Dewdney} et~al.}{2009}]{Dewdney2009}
{Dewdney} P.~E.,  {Hall} P.~J.,  {Schilizzi} R.~T.,   {Lazio} T.~J.~L.~W.,  2009, \mn@doi [IEEE Proc] {10.1109/JPROC.2009.2021005}, \href {https://ui.adsabs.harvard.edu/abs/2009IEEEP..97.1482D} {97, 1482}

\bibitem[\protect\citeauthoryear{{Faucher-Gigu{\`e}re} \& {Kaspi}}{{Faucher-Gigu{\`e}re} \& {Kaspi}}{2006}]{Faucher2006}
{Faucher-Gigu{\`e}re} C.-A.,  {Kaspi} V.~M.,  2006, \mn@doi [\apj] {10.1086/501516}, \href {https://ui.adsabs.harvard.edu/abs/2006ApJ...643..332F} {643, 332}

\bibitem[\protect\citeauthoryear{{Geyer} et~al.,}{{Geyer} et~al.}{2017}]{Geyer2017}
{Geyer} M.,  et~al., 2017, \mn@doi [\mnras] {10.1093/mnras/stx1151}, \href {https://ui.adsabs.harvard.edu/abs/2017MNRAS.470.2659G} {470, 2659}

\bibitem[\protect\citeauthoryear{{Gil}, {Gronkowski}  \& {Rudnicki}}{{Gil} et~al.}{1984}]{Gil1984}
{Gil} J.,  {Gronkowski} P.,   {Rudnicki} W.,  1984, \aap, \href {https://ui.adsabs.harvard.edu/abs/1984A&A...132..312G} {132, 312}

\bibitem[\protect\citeauthoryear{{Gong} et~al.,}{{Gong} et~al.}{2020}]{Gong2020}
{Gong} H.,  et~al., 2020, \mn@doi [Scientia Sinica Physica, Mechanica \& Astronomica] {10.1360/SSPMA-2020-0064}, \href {https://ui.adsabs.harvard.edu/abs/2020SSPMA..50j9501G} {50, 109501}

\bibitem[\protect\citeauthoryear{{He} \& {Shi}}{{He} \& {Shi}}{2024}]{He2024}
{He} Q.,  {Shi} X.,  2024, \mn@doi [\mnras] {10.1093/mnras/stad3561}, \href {https://ui.adsabs.harvard.edu/abs/2024MNRAS.527.5183H} {527, 5183}

\bibitem[\protect\citeauthoryear{{Hewish}, {Bell}, {Pilkington}, {Scott}  \& {Collins}}{{Hewish} et~al.}{1968}]{Hewish1968}
{Hewish} A.,  {Bell} S.~J.,  {Pilkington} J.~D.~H.,  {Scott} P.~F.,   {Collins} R.~A.,  1968, \mn@doi [\nat] {10.1038/217709a0}, \href {https://ui.adsabs.harvard.edu/abs/1968Natur.217..709H} {217, 709}

\bibitem[\protect\citeauthoryear{{Hobbs}, {Lorimer}, {Lyne}  \& {Kramer}}{{Hobbs} et~al.}{2005}]{Hobbs2005}
{Hobbs} G.,  {Lorimer} D.~R.,  {Lyne} A.~G.,   {Kramer} M.,  2005, \mn@doi [\mnras] {10.1111/j.1365-2966.2005.09087.x}, \href {https://ui.adsabs.harvard.edu/abs/2005MNRAS.360..974H} {360, 974}

\bibitem[\protect\citeauthoryear{{Hurley-Walker} et~al.,}{{Hurley-Walker} et~al.}{2023}]{Hurley-Walker2023}
{Hurley-Walker} N.,  et~al., 2023, \mn@doi [\nat] {10.1038/s41586-023-06202-5}, \href {https://ui.adsabs.harvard.edu/abs/2023Natur.619..487H} {619, 487}

\bibitem[\protect\citeauthoryear{{Hurley-Walker} et~al.,}{{Hurley-Walker} et~al.}{2024}]{Hurley-Walker2024}
{Hurley-Walker} N.,  et~al., 2024, \mn@doi [arXiv e-prints] {10.48550/arXiv.2408.15757}, \href {https://ui.adsabs.harvard.edu/abs/2024arXiv240815757H} {p. arXiv:2408.15757}

\bibitem[\protect\citeauthoryear{{Izvekova}, {Kuzmin}, {Malofeev}  \& {Shitov}}{{Izvekova} et~al.}{1981}]{Izvekova1981}
{Izvekova} V.~A.,  {Kuzmin} A.~D.,  {Malofeev} V.~M.,   {Shitov} I.~P.,  1981, \mn@doi [\apss] {10.1007/BF00654022}, \href {https://ui.adsabs.harvard.edu/abs/1981Ap&SS..78...45I} {78, 45}

\bibitem[\protect\citeauthoryear{{Janagal}, {Chakraborty}, {Bhat}, {McSweeney}  \& {Sett}}{{Janagal} et~al.}{2023}]{Janagal2023}
{Janagal} P.,  {Chakraborty} M.,  {Bhat} N.~D.~R.,  {McSweeney} S.~J.,   {Sett} S.,  2023, \mn@doi [\mnras] {10.1093/mnras/stad1797}, \href {https://ui.adsabs.harvard.edu/abs/2023MNRAS.523.5934J} {523, 5934}

\bibitem[\protect\citeauthoryear{{Jankowski}, {van Straten}, {Keane}, {Bailes}, {Barr}, {Johnston}  \& {Kerr}}{{Jankowski} et~al.}{2018}]{Jankowski2018}
{Jankowski} F.,  {van Straten} W.,  {Keane} E.~F.,  {Bailes} M.,  {Barr} E.~D.,  {Johnston} S.,   {Kerr} M.,  2018, \mn@doi [\mnras] {10.1093/mnras/stx2476}, \href {https://ui.adsabs.harvard.edu/abs/2018MNRAS.473.4436J} {473, 4436}

\bibitem[\protect\citeauthoryear{{Johnston} \& {Karastergiou}}{{Johnston} \& {Karastergiou}}{2019}]{Johnston2019}
{Johnston} S.,  {Karastergiou} A.,  2019, \mn@doi [\mnras] {10.1093/mnras/stz400}, \href {https://ui.adsabs.harvard.edu/abs/2019MNRAS.485..640J} {485, 640}

\bibitem[\protect\citeauthoryear{{Keane} et~al.,}{{Keane} et~al.}{2015}]{Keane2015}
{Keane} E.,  et~al., 2015, in Advancing Astrophysics with the Square Kilometre Array (AASKA14). p.~40 (\mn@eprint {arXiv} {1501.00056}), \mn@doi{10.22323/1.215.0040}

\bibitem[\protect\citeauthoryear{{Kijak} \& {Gil}}{{Kijak} \& {Gil}}{1998}]{Kijak1998}
{Kijak} J.,  {Gil} J.,  1998, \mn@doi [\mnras] {10.1046/j.1365-8711.1998.01832.x}, \href {https://ui.adsabs.harvard.edu/abs/1998MNRAS.299..855K} {299, 855}

\bibitem[\protect\citeauthoryear{{Kijak} \& {Gil}}{{Kijak} \& {Gil}}{2003}]{Kijak2003}
{Kijak} J.,  {Gil} J.,  2003, \mn@doi [\aap] {10.1051/0004-6361:20021583}, \href {https://ui.adsabs.harvard.edu/abs/2003A&A...397..969K} {397, 969}

\bibitem[\protect\citeauthoryear{{Kijak}, {Basu}, {Lewandowski}  \& {Ro{\.z}ko}}{{Kijak} et~al.}{2021}]{Kijak2021}
{Kijak} J.,  {Basu} R.,  {Lewandowski} W.,   {Ro{\.z}ko} K.,  2021, \mn@doi [\apj] {10.3847/1538-4357/ac3082}, \href {https://ui.adsabs.harvard.edu/abs/2021ApJ...923..211K} {923, 211}

\bibitem[\protect\citeauthoryear{{Kirsten}, {Bhat}, {Meyers}, {Macquart}, {Tremblay}  \& {Ord}}{{Kirsten} et~al.}{2019}]{Kirsten2019}
{Kirsten} F.,  {Bhat} N.~D.~R.,  {Meyers} B.~W.,  {Macquart} J.~P.,  {Tremblay} S.~E.,   {Ord} S.~M.,  2019, \mn@doi [\apj] {10.3847/1538-4357/ab0c05}, \href {https://ui.adsabs.harvard.edu/abs/2019ApJ...874..179K} {874, 179}

\bibitem[\protect\citeauthoryear{{Koribalski}, {Johnston}, {Weisberg}  \& {Wilson}}{{Koribalski} et~al.}{1995}]{Koribalski1995}
{Koribalski} B.,  {Johnston} S.,  {Weisberg} J.~M.,   {Wilson} W.,  1995, \mn@doi [\apj] {10.1086/175397}, \href {https://ui.adsabs.harvard.edu/abs/1995ApJ...441..756K} {441, 756}

\bibitem[\protect\citeauthoryear{{Krishnakumar}, {Mitra}, {Naidu}, {Joshi}  \& {Manoharan}}{{Krishnakumar} et~al.}{2015}]{Krishnakumar2015}
{Krishnakumar} M.~A.,  {Mitra} D.,  {Naidu} A.,  {Joshi} B.~C.,   {Manoharan} P.~K.,  2015, \mn@doi [\apj] {10.1088/0004-637X/804/1/23}, \href {https://ui.adsabs.harvard.edu/abs/2015ApJ...804...23K} {804, 23}

\bibitem[\protect\citeauthoryear{{Kuniyoshi}, {Verbiest}, {Lee}, {Adebahr}, {Kramer}  \& {Noutsos}}{{Kuniyoshi} et~al.}{2015}]{Kuniyoshi2015}
{Kuniyoshi} M.,  {Verbiest} J.~P.~W.,  {Lee} K.~J.,  {Adebahr} B.,  {Kramer} M.,   {Noutsos} A.,  2015, \mn@doi [\mnras] {10.1093/mnras/stv1604}, \href {https://ui.adsabs.harvard.edu/abs/2015MNRAS.453..828K} {453, 828}

\bibitem[\protect\citeauthoryear{{Lee} et~al.,}{{Lee} et~al.}{2025}]{Lee2025}
{Lee} C.~P.,  et~al., 2025, \mn@doi [arXiv e-prints] {10.48550/arXiv.2508.10330}, \href {https://ui.adsabs.harvard.edu/abs/2025arXiv250810330L} {p. arXiv:2508.10330}

\bibitem[\protect\citeauthoryear{{Lewandowski}, {Dembska}, {Kijak}  \& {Kowali{\'n}ska}}{{Lewandowski} et~al.}{2013}]{Lewandowski2013}
{Lewandowski} W.,  {Dembska} M.,  {Kijak} J.,   {Kowali{\'n}ska} M.,  2013, \mn@doi [\mnras] {10.1093/mnras/stt989}, \href {https://ui.adsabs.harvard.edu/abs/2013MNRAS.434...69L} {434, 69}

\bibitem[\protect\citeauthoryear{{Li} \& {Gao}}{{Li} \& {Gao}}{2023}]{Li2023}
{Li} B.-P.,  {Gao} Z.-F.,  2023, \mn@doi [Astron. Nachr.] {10.1002/asna.20220111}, \href {https://ui.adsabs.harvard.edu/abs/2023AN....34420111L} {344, e20220111}

\bibitem[\protect\citeauthoryear{{Li}, {Ma}  \& {Gao}}{{Li} et~al.}{2024}]{Li2024}
{Li} B.-P.,  {Ma} W.-Q.,   {Gao} Z.-F.,  2024, \mn@doi [Astron. Nachr.] {10.1002/asna.20230167}, \href {https://ui.adsabs.harvard.edu/abs/2024AN....34530167L} {345, e20230167}

\bibitem[\protect\citeauthoryear{{L{\"o}hmer}, {Kramer}, {Mitra}, {Lorimer}  \& {Lyne}}{{L{\"o}hmer} et~al.}{2001}]{Lohmer2001}
{L{\"o}hmer} O.,  {Kramer} M.,  {Mitra} D.,  {Lorimer} D.~R.,   {Lyne} A.~G.,  2001, \mn@doi [\apjl] {10.1086/338324}, \href {https://ui.adsabs.harvard.edu/abs/2001ApJ...562L.157L} {562, L157}

\bibitem[\protect\citeauthoryear{{L{\"o}hmer}, {Mitra}, {Gupta}, {Kramer}  \& {Ahuja}}{{L{\"o}hmer} et~al.}{2004}]{Lohmer2004}
{L{\"o}hmer} O.,  {Mitra} D.,  {Gupta} Y.,  {Kramer} M.,   {Ahuja} A.,  2004, \mn@doi [\aap] {10.1051/0004-6361:20035881}, \href {https://ui.adsabs.harvard.edu/abs/2004A&A...425..569L} {425, 569}

\bibitem[\protect\citeauthoryear{{Lorimer} \& {Kramer}}{{Lorimer} \& {Kramer}}{2004}]{Lorimer2004}
{Lorimer} D.~R.,  {Kramer} M.,  2004, {Handbook of Pulsar Astronomy}.
Cambridge University Press, Cambridge

\bibitem[\protect\citeauthoryear{{Lynch} et~al.,}{{Lynch} et~al.}{2018}]{Lynch2018}
{Lynch} R.~S.,  et~al., 2018, in American Astronomical Society Meeting Abstracts \#231. p. 243.13

\bibitem[\protect\citeauthoryear{{Lynch} et~al.,}{{Lynch} et~al.}{2021}]{Lynch2021}
{Lynch} R.,  et~al., 2021, in American Astronomical Society Meeting Abstracts. p. 345.01

\bibitem[\protect\citeauthoryear{{Lyne}, {Manchester}  \& {Taylor}}{{Lyne} et~al.}{1985}]{Lyne1985}
{Lyne} A.~G.,  {Manchester} R.~N.,   {Taylor} J.~H.,  1985, \mn@doi [\mnras] {10.1093/mnras/213.3.613}, \href {https://ui.adsabs.harvard.edu/abs/1985MNRAS.213..613L} {213, 613}

\bibitem[\protect\citeauthoryear{{Manchester}, {Hobbs}, {Teoh}  \& {Hobbs}}{{Manchester} et~al.}{2005}]{Manchester2005}
{Manchester} R.~N.,  {Hobbs} G.~B.,  {Teoh} A.,   {Hobbs} M.,  2005, \mn@doi [\aj] {10.1086/428488}, \href {https://ui.adsabs.harvard.edu/abs/2005AJ....129.1993M} {129, 1993}

\bibitem[\protect\citeauthoryear{{Mantovanini}, {Hurley-Walker}, {Anderson}, {Ross}, {Duchesne}  \& {Galvin}}{{Mantovanini} et~al.}{2025}]{Mantovanini2025}
{Mantovanini} S.,  {Hurley-Walker} N.,  {Anderson} G.,  {Ross} K.,  {Duchesne} S.~W.,   {Galvin} T.~J.,  2025, \mn@doi [\mnras] {10.1093/mnras/staf1429}, \href {https://ui.adsabs.harvard.edu/abs/2025MNRAS.tmp.1383M} {}

\bibitem[\protect\citeauthoryear{{Martinez} et~al.,}{{Martinez} et~al.}{2019}]{Martinez2019}
{Martinez} J.~G.,  et~al., 2019, \mn@doi [\apj] {10.3847/1538-4357/ab2877}, \href {https://ui.adsabs.harvard.edu/abs/2019ApJ...881..166M} {881, 166}

\bibitem[\protect\citeauthoryear{{McEwen} et~al.,}{{McEwen} et~al.}{2020}]{McEwen2020}
{McEwen} A.~E.,  et~al., 2020, \mn@doi [\apj] {10.3847/1538-4357/ab75e2}, \href {https://ui.adsabs.harvard.edu/abs/2020ApJ...892...76M} {892, 76}

\bibitem[\protect\citeauthoryear{{Meyers} et~al.,}{{Meyers} et~al.}{2017}]{Meyers2017}
{Meyers} B.~W.,  et~al., 2017, \mn@doi [\apj] {10.3847/1538-4357/aa8bba}, \href {https://ui.adsabs.harvard.edu/abs/2017ApJ...851...20M} {851, 20}

\bibitem[\protect\citeauthoryear{{Michilli} et~al.,}{{Michilli} et~al.}{2020}]{Michilli2020}
{Michilli} D.,  et~al., 2020, \mn@doi [\mnras] {10.1093/mnras/stz2997}, \href {https://ui.adsabs.harvard.edu/abs/2020MNRAS.491..725M} {491, 725}

\bibitem[\protect\citeauthoryear{Montgomery, Peck  \& Vining}{Montgomery et~al.}{2012}]{montgomery2012introduction}
Montgomery D.~C.,  Peck E.~A.,   Vining G.~G.,  2012, Introduction to Linear Regression Analysis, 5th edn.
Wiley, Hoboken, NJ

\bibitem[\protect\citeauthoryear{{Ochelkov} \& {Usov}}{{Ochelkov} \& {Usov}}{1984}]{Ochelkov1984}
{Ochelkov} I.~P.,  {Usov} V.~V.,  1984, \mn@doi [\nat] {10.1038/309332a0}, \href {https://ui.adsabs.harvard.edu/abs/1984Natur.309..332O} {309, 332}

\bibitem[\protect\citeauthoryear{{Oronsaye} et~al.,}{{Oronsaye} et~al.}{2015}]{Oronsaye2015}
{Oronsaye} S.~I.,  et~al., 2015, \mn@doi [\apj] {10.1088/0004-637X/809/1/51}, \href {https://ui.adsabs.harvard.edu/abs/2015ApJ...809...51O} {809, 51}

\bibitem[\protect\citeauthoryear{{Pilia} et~al.,}{{Pilia} et~al.}{2016}]{Pilia2016}
{Pilia} M.,  et~al., 2016, \mn@doi [\aap] {10.1051/0004-6361/201425196}, \href {https://ui.adsabs.harvard.edu/abs/2016A&A...586A..92P} {586, A92}

\bibitem[\protect\citeauthoryear{{Posselt} et~al.,}{{Posselt} et~al.}{2021}]{Posselt2021}
{Posselt} B.,  et~al., 2021, \mn@doi [\mnras] {10.1093/mnras/stab2775}, \href {https://ui.adsabs.harvard.edu/abs/2021MNRAS.508.4249P} {508, 4249}

\bibitem[\protect\citeauthoryear{{Prabu} et~al.,}{{Prabu} et~al.}{2015}]{Prabu2015}
{Prabu} T.,  et~al., 2015, \mn@doi [Experimental Astronomy] {10.1007/s10686-015-9444-3}, \href {https://ui.adsabs.harvard.edu/abs/2015ExA....39...73P} {39, 73}

\bibitem[\protect\citeauthoryear{{Rankin}}{{Rankin}}{1990}]{Rankin1990}
{Rankin} J.~M.,  1990, \mn@doi [\apj] {10.1086/168530}, \href {https://ui.adsabs.harvard.edu/abs/1990ApJ...352..247R} {352, 247}

\bibitem[\protect\citeauthoryear{{Rankin}}{{Rankin}}{1993}]{Rankin1993}
{Rankin} J.~M.,  1993, \mn@doi [\apj] {10.1086/172361}, \href {https://ui.adsabs.harvard.edu/abs/1993ApJ...405..285R} {405, 285}

\bibitem[\protect\citeauthoryear{{Ransom}}{{Ransom}}{2001}]{Ransom2001}
{Ransom} S.~M.,  2001, PhD thesis, Harvard University, Massachusetts

\bibitem[\protect\citeauthoryear{{Ransom}, {Eikenberry}  \& {Middleditch}}{{Ransom} et~al.}{2002}]{Ransom2002}
{Ransom} S.~M.,  {Eikenberry} S.~S.,   {Middleditch} J.,  2002, \mn@doi [\aj] {10.1086/342285}, \href {https://ui.adsabs.harvard.edu/abs/2002AJ....124.1788R} {124, 1788}

\bibitem[\protect\citeauthoryear{{Rickett}}{{Rickett}}{1977}]{Rickett1977}
{Rickett} B.~J.,  1977, \mn@doi [\araa] {10.1146/annurev.aa.15.090177.002403}, \href {https://ui.adsabs.harvard.edu/abs/1977ARA&A..15..479R} {15, 479}

\bibitem[\protect\citeauthoryear{{Rickett}}{{Rickett}}{1990}]{Rickett1990}
{Rickett} B.~J.,  1990, \mn@doi [\araa] {10.1146/annurev.aa.28.090190.003021}, \href {https://ui.adsabs.harvard.edu/abs/1990ARA&A..28..561R} {28, 561}

\bibitem[\protect\citeauthoryear{{Romani}, {Narayan}  \& {Blandford}}{{Romani} et~al.}{1986}]{Romani1986}
{Romani} R.~W.,  {Narayan} R.,   {Blandford} R.,  1986, \mn@doi [\mnras] {10.1093/mnras/220.1.19}, \href {https://ui.adsabs.harvard.edu/abs/1986MNRAS.220...19R} {220, 19}

\bibitem[\protect\citeauthoryear{{Sanidas} et~al.,}{{Sanidas} et~al.}{2019}]{Sanidas2019}
{Sanidas} S.,  et~al., 2019, \mn@doi [\aap] {10.1051/0004-6361/201935609}, \href {https://ui.adsabs.harvard.edu/abs/2019A&A...626A.104S} {626, A104}

\bibitem[\protect\citeauthoryear{{Shapiro} \& {Teukolsky}}{{Shapiro} \& {Teukolsky}}{1983}]{Shapiro1983}
{Shapiro} S.~L.,  {Teukolsky} S.~A.,  1983, {Black holes, white dwarfs and neutron stars. The physics of compact objects}, \mn@doi{10.1002/9783527617661.
}

\bibitem[\protect\citeauthoryear{{Sieber}}{{Sieber}}{1973}]{Sieber1973}
{Sieber} W.,  1973, \aap, \href {https://ui.adsabs.harvard.edu/abs/1973A&A....28..237S} {28, 237}

\bibitem[\protect\citeauthoryear{{Singh}, {Roy}, {Panda}, {Bhattacharyya}, {Morello}, {Stappers}, {Ray}  \& {McLaughlin}}{{Singh} et~al.}{2022}]{Singh2022}
{Singh} S.,  {Roy} J.,  {Panda} U.,  {Bhattacharyya} B.,  {Morello} V.,  {Stappers} B.~W.,  {Ray} P.~S.,   {McLaughlin} M.~A.,  2022, \mn@doi [\apj] {10.3847/1538-4357/ac7b91}, \href {https://ui.adsabs.harvard.edu/abs/2022ApJ...934..138S} {934, 138}

\bibitem[\protect\citeauthoryear{{Singh}, {Roy}, {Bhattachryya}, {Sharma}  \& {Panda}}{{Singh} et~al.}{2023}]{Singh2023}
{Singh} S.,  {Roy} J.,  {Bhattachryya} B.,  {Sharma} S.~S.,   {Panda} U.,  2023, in 2023 XXXVth General Assembly and Scientific Symposium of the International Union of Radio Science (URSI GASS. p.~328, \mn@doi{10.23919/URSIGASS57860.2023.10265649}

\bibitem[\protect\citeauthoryear{{Sokolowski} et~al.,}{{Sokolowski} et~al.}{2017}]{Sokolowski2017}
{Sokolowski} M.,  et~al., 2017, \mn@doi [PASA] {10.1017/pasa.2017.54}, \href {https://ui.adsabs.harvard.edu/abs/2017PASA...34...62S} {34, e062}

\bibitem[\protect\citeauthoryear{{Stovall} et~al.,}{{Stovall} et~al.}{2014}]{Stovall2014}
{Stovall} K.,  et~al., 2014, \mn@doi [\apj] {10.1088/0004-637X/791/1/67}, \href {https://ui.adsabs.harvard.edu/abs/2014ApJ...791...67S} {791, 67}

\bibitem[\protect\citeauthoryear{{Sunder}, {Roy}, {Kudale}, {Bhattacharyya}, {Behera}  \& {Singh}}{{Sunder} et~al.}{2023}]{Sunder2023}
{Sunder} S.,  {Roy} J.,  {Kudale} S.,  {Bhattacharyya} B.,  {Behera} A.~K.,   {Singh} S.,  2023, \mn@doi [arXiv e-prints] {10.48550/arXiv.2302.13363}, \href {https://ui.adsabs.harvard.edu/abs/2023arXiv230213363S} {p. arXiv:2302.13363}

\bibitem[\protect\citeauthoryear{{Swainston}, {Lee}, {McSweeney}  \& {Bhat}}{{Swainston} et~al.}{2022}]{Swainston2022}
{Swainston} N.~A.,  {Lee} C.~P.,  {McSweeney} S.~J.,   {Bhat} N.~D.~R.,  2022, \mn@doi [PASA] {10.1017/pasa.2022.52}, \href {https://ui.adsabs.harvard.edu/abs/2022PASA...39...56S} {39, e056}

\bibitem[\protect\citeauthoryear{{Tan} et~al.,}{{Tan} et~al.}{2018}]{Tan2018}
{Tan} C.~M.,  et~al., 2018, \mn@doi [\apj] {10.3847/1538-4357/aade88}, \href {https://ui.adsabs.harvard.edu/abs/2018ApJ...866...54T} {866, 54}

\bibitem[\protect\citeauthoryear{{Tan} et~al.,}{{Tan} et~al.}{2020}]{Tan2020}
{Tan} C.~M.,  et~al., 2020, \mn@doi [\mnras] {10.1093/mnras/staa113}, \href {https://ui.adsabs.harvard.edu/abs/2020MNRAS.492.5878T} {492, 5878}

\bibitem[\protect\citeauthoryear{{Taylor} \& {Cordes}}{{Taylor} \& {Cordes}}{1993}]{Taylor1993}
{Taylor} J.~H.,  {Cordes} J.~M.,  1993, \mn@doi [\apj] {10.1086/172870}, \href {https://ui.adsabs.harvard.edu/abs/1993ApJ...411..674T} {411, 674}

\bibitem[\protect\citeauthoryear{{Tingay} et~al.,}{{Tingay} et~al.}{2013}]{Tingay2013}
{Tingay} S.~J.,  et~al., 2013, \mn@doi [PASA] {10.1017/pasa.2012.007}, \href {https://ui.adsabs.harvard.edu/abs/2013PASA...30....7T} {30, e007}

\bibitem[\protect\citeauthoryear{{Tremblay} et~al.,}{{Tremblay} et~al.}{2015}]{Tremblay2015}
{Tremblay} S.~E.,  et~al., 2015, \mn@doi [PASA] {10.1017/pasa.2015.6}, \href {https://ui.adsabs.harvard.edu/abs/2015PASA...32....5T} {32, e005}

\bibitem[\protect\citeauthoryear{{Tyul'bashev}, {Tyul'basheva}  \& {Kitaeva}}{{Tyul'bashev} et~al.}{2022}]{Tyul'bashev2022}
{Tyul'bashev} S.~A.,  {Tyul'basheva} G.~E.,   {Kitaeva} M.~A.,  2022, in The Multifaceted Universe: Theory and Observations - 2000. p.~43 (\mn@eprint {arXiv} {2208.04578}), \mn@doi{10.48550/arXiv.2208.04578}

\bibitem[\protect\citeauthoryear{{Tyul'bashev}, {Tyul'basheva}, {Kitaeva}, {Ovchinnikov}, {Oreshko}  \& {Logvinenko}}{{Tyul'bashev} et~al.}{2024}]{Tyul'bashev2024}
{Tyul'bashev} S.~A.,  {Tyul'basheva} G.~E.,  {Kitaeva} M.~A.,  {Ovchinnikov} I.~L.,  {Oreshko} V.~V.,   {Logvinenko} S.~V.,  2024, \mn@doi [\mnras] {10.1093/mnras/stae070}, \href {https://ui.adsabs.harvard.edu/abs/2024MNRAS.528.2220T} {528, 2220}

\bibitem[\protect\citeauthoryear{{Wang}, {Wu}, {Manchester}, {Zhang}, {Yusup}  \& {Zhang}}{{Wang} et~al.}{2001}]{Wang2001}
{Wang} N.,  {Wu} X.-J.,  {Manchester} R.~N.,  {Zhang} J.,  {Yusup} A.,   {Zhang} H.-B.,  2001, \mn@doi [\cjaa] {10.1088/1009-9271/1/5/421}, \href {https://ui.adsabs.harvard.edu/abs/2001ChJAA...1..421W} {1, 421}

\bibitem[\protect\citeauthoryear{{Wang}, {Manchester}, {Johnston}, {Rickett}, {Zhang}, {Yusup}  \& {Chen}}{{Wang} et~al.}{2005}]{Wang2005}
{Wang} N.,  {Manchester} R.~N.,  {Johnston} S.,  {Rickett} B.,  {Zhang} J.,  {Yusup} A.,   {Chen} M.,  2005, \mn@doi [\mnras] {10.1111/j.1365-2966.2005.08798.x}, \href {https://ui.adsabs.harvard.edu/abs/2005MNRAS.358..270W} {358, 270}

\bibitem[\protect\citeauthoryear{{Wei}, {Zhang}, {Zhang}, {Yu}, {LIN}  \& {An}}{{Wei} et~al.}{2023}]{Wei2023}
{Wei} J.,  {Zhang} C.,  {Zhang} Z.,  {Yu} T.,  {LIN} J.,   {An} T.,  2023, \mn@doi [Sci. China Phys. Mech. Astron] {10.1360/SSPMA-2022-0264}, \href {https://www.sciengine.com/SSPMA/doi/10.1360/SSPMA-2022-0264;JSESSIONID=44911a17-3231-424e-97ab-61d5f8942cf5} {53, 229506}

\bibitem[\protect\citeauthoryear{{Williamson}}{{Williamson}}{1972}]{Williamson1972}
{Williamson} I.~P.,  1972, \mn@doi [\mnras] {10.1093/mnras/157.1.55}, \href {https://ui.adsabs.harvard.edu/abs/1972MNRAS.157...55W} {157, 55}

\bibitem[\protect\citeauthoryear{{Williamson}}{{Williamson}}{1973}]{Williamson1973}
{Williamson} I.~P.,  1973, \mn@doi [\mnras] {10.1093/mnras/163.4.345}, \href {https://ui.adsabs.harvard.edu/abs/1973MNRAS.163..345W} {163, 345}

\bibitem[\protect\citeauthoryear{{Xue} et~al.,}{{Xue} et~al.}{2017}]{Xue2017}
{Xue} M.,  et~al., 2017, \mn@doi [PASA] {10.1017/pasa.2017.66}, \href {https://ui.adsabs.harvard.edu/abs/2017PASA...34...70X} {34, e070}

\bibitem[\protect\citeauthoryear{{Yao}, {Manchester}  \& {Wang}}{{Yao} et~al.}{2017}]{Yao2017}
{Yao} J.~M.,  {Manchester} R.~N.,   {Wang} N.,  2017, \mn@doi [\apj] {10.3847/1538-4357/835/1/29}, \href {https://ui.adsabs.harvard.edu/abs/2017ApJ...835...29Y} {835, 29}

\bibitem[\protect\citeauthoryear{{Zhou}, {Kurban}, {Liu}, {Wang}  \& {Yuan}}{{Zhou} et~al.}{2025}]{Zhou2025}
{Zhou} X.,  {Kurban} A.,  {Liu} W.-T.,  {Wang} N.,   {Yuan} Y.-J.,  2025, \mn@doi [\apj] {10.3847/1538-4357/add268}, \href {https://ui.adsabs.harvard.edu/abs/2025ApJ...986...98Z} {986, 98}

\bibitem[\protect\citeauthoryear{{van der Wateren} et~al.,}{{van der Wateren} et~al.}{2023}]{Wateren2023}
{van der Wateren} E.,  et~al., 2023, \mn@doi [\aap] {10.1051/0004-6361/202245122}, \href {https://ui.adsabs.harvard.edu/abs/2023A&A...669A.160V} {669, A160}

\makeatother
\end{thebibliography}

\appendix            
\begin{table*}\section{Observation Details}\label{appendix:obs_table} 
\caption{Summary of the 48 MWA-VCS drift-scan observations used in this study}
\begin{tabular}{ccccccccc}
\hline\noalign{\smallskip}
Obs. ID & RA (J2000) & DEC (J2000) & MJD & $T_{\text{sys}}$ (K) & $G_{\text{peak}}^{\ast}$ (K/Jy) & Tiles (N) & $t_{\text{int}}$ (s) \\
\hline\noalign{\smallskip}
1088850560	&	13:20:07.2000	&	-26:37:12.0000	&	56846	&	263.93	&	0.0271	&	125	&	3535	\\
1090852664	&	18:59:33.6000	&	-47:33:36.0000	&	56869	&	398.28	&	0.0236	&	128	&	2703	\\
1091309464	&	02:14:24.0000	&	-55:04:48.0000	&	56874	&	182.98	&	0.0207	&	128	&	2922	\\
1091793216	&	16:58:21.6000	&	-47:30:36.0000	&	56880	&	972.39	&	0.0237	&	128	&	3848	\\
1095506112	&	19:09:50.4000	&	+18:36:00.0000	&	56923	&	469.91	&	0.0124	&	126	&	3947	\\
1097404000	&	19:47:48.4066	&	-05:54:36.3530	&	56945	&	410.57	&	0.0221	&	112	&	3848	\\
1098439592	&	20:14:38.4000	&	-47:34:48.0000	&	56957	&	319.01	&	0.0236	&	128	&	3852	\\
1099414416	&	05:34:31.2000	&	+22:00:36.0000	&	56968	&	238.24	&	0.0086	&	128	&	1164	\\
1100288192	&	05:35:33.6000	&	-53:34:48.0000	&	56978	&	187.78	&	0.019	&	128	&	4837	\\
1101491208	&	06:14:36.0000	&	-26:41:24.0000	&	56992	&	182.63	&	0.0225	&	86	&	4859	\\
1101925816	&	07:17:51.6658	&	-19:51:56.3599	&	56997	&	233.52	&	0.022	&	85	&	1744	\\
1107478712	&	17:59:14.4000	&	-26:41:60.0000	&	57062	&	1178.48	&	0.0214	&	78	&	2464	\\
1115381072	&	10:10:08.1228	&	+10:39:45.9377	&	57153	&	168.04	&	0.0152	&	126	&	4868	\\
1116090392	&	17:08:12.1065	&	-12:46:40.9635	&	57161	&	566.82	&	0.025	&	125	&	3668	\\
1116787952	&	18:22:10.7591	&	-40:11:11.5443	&	57169	&	676.17	&	0.025	&	125	&	4800	\\
1117899328	&	16:31:52.5383	&	+09:40:52.9090	&	57182	&	373.39	&	0.0166	&	123	&	4868	\\
1118168248	&	21:21:06.6576	&	-09:48:32.3615	&	57185	&	258.9	&	0.0191	&	125	&	4864	\\
1121173352	&	14:53:47.2999	&	-53:30:44.5651	&	57220	&	634.96	&	0.0188	&	125	&	4914	\\
1123367368	&	04:15:55.3517	&	-47:17:29.5113	&	57245	&	163.24	&	0.0231	&	126	&	2549	\\
1127329112	&	03:37:58.0239	&	-33:20:45.2485	&	57291	&	149.87	&	0.0263	&	125	&	3716	\\
1128772184	&	21:02:15.3550	&	+09:35:02.5317	&	57308	&	253.23	&	0.0168	&	125	&	3750	\\
1129464688	&	21:55:17.5216	&	-55:05:09.9566	&	57316	&	222.68	&	0.0205	&	125	&	5075	\\
1131415232	&	14:39:27.4317	&	-71:09:36.8236	&	57339	&	373.14	&	0.012	&	125	&	594	\\
1131957328	&	20:13:28.1755	&	+29:19:49.3530	&	57345	&	335.74	&	0.0073	&	125	&	593	\\
1132030232	&	16:31:40.6405	&	-05:50:21.6172	&	57346	&	388.09	&	0.0234	&	125	&	718	\\
1133329792	&	18:30:10.1514	&	-26:43:02.5351	&	57361	&	898.69	&	0.026	&	115	&	3714	\\
1133775752	&	22:43:19.8013	&	-05:57:11.7639	&	57366	&	192.24	&	0.0227	&	118	&	4914	\\
1137236608	&	00:15:14.6823	&	-07:47:25.2123	&	57406	&	200.67	&	0.0147	&	118	&	2514	\\
1139239952	&	08:42:52.0325	&	-40:21:26.0914	&	57429	&	314.56	&	0.0247	&	118	&	1274	\\
1139324488	&	08:15:23.3922	&	+09:42:18.1324	&	57430	&	182.34	&	0.0163	&	118	&	2714	\\
1140972392	&	11:15:40.7853	&	+18:43:53.9555	&	57449	&	235.25	&	0.012	&	118	&	2514	\\
1145367872	&	11:34:17.2956	&	-33:25:06.9706	&	57500	&	194.64	&	0.0259	&	118	&	3613	\\
1148063920	&	19:29:55.1220	&	+19:43:45.0742	&	57531	&	435.98	&	0.0103	&	114	&	2594	\\
1151251888	&	18:29:41.8642	&	+01:35:26.7556	&	57568	&	806.47	&	0.0197	&	116	&	434	\\
1152636328	&	20:06:47.3537	&	-13:02:19.5415	&	57584	&	309.78	&	0.0238	&	109	&	4794	\\
1163853320	&	00:28:40.0738	&	-63:08:47.3443	&	57714	&	193.66	&	0.0168	&	125	&	5234	\\
1164474312	&	05:26:48.0933	&	-05:52:49.9614	&	57721	&	211.77	&	0.0233	&	124	&	1216	\\
1169379256	&	03:39:47.8905	&	-26:45:07.0030	&	57778	&	149.49	&	0.0254	&	110	&	3712	\\
1171113560	&	06:43:46.3844	&	+09:40:22.5968	&	57798	&	240.39	&	0.0154	&	106	&	1909	\\
1173020416	&	09:52:02.2051	&	-54:55:41.2600	&	57820	&	305.96	&	0.0174	&	90	&	3712	\\
1173793160	&	09:06:09.8861	&	-12:54:49.4659	&	57829	&	169.17	&	0.0221	&	94	&	3712	\\
1177428616	&	13:42:43.3256	&	-47:27:03.0101	&	57871	&	428.52	&	0.0208	&	99	&	4914	\\
1177766416	&	11:48:20.6620	&	-19:47:46.2209	&	57875	&	185.29	&	0.0235	&	97	&	3714	\\
1178380936	&	14:58:23.0164	&	+01:40:24.5190	&	57882	&	281.11	&	0.0181	&	98	&	2514	\\
1178471536	&	16:12:07.3945	&	-47:29:47.7599	&	57883	&	940.89	&	0.0208	&	99	&	1314	\\
1178553136	&	14:56:16.7717	&	+09:43:10.6434	&	57884	&	284.11	&	0.0148	&	98	&	2514	\\
1185884600	&	17:01:17.8120	&	-54:59:14.3606	&	57969	&	731.28	&	0.0201	&	121	&	3714	\\
1186248016	&	22:15:24.4262	&	+18:33:05.7255	&	57973	&	224.25	&	0.012	&	118	&	3712	\\
\hline
\end{tabular}
\vspace{2mm}

\raggedright
\footnotesize{\textbf{Note.} $^{\ast}$ $G_{\text{peak}}$ denotes the peak gain in incoherent-sum mode for each observation. Only two entries are shown here as an example; the full table includes all 48 observations.}
\end{table*}

\clearpage

\renewcommand{\arraystretch}{1.2} 
\setlength{\tabcolsep}{4pt}      
\onecolumn
\section{Cataloged Parameters of the 80 Detected Pulsars}

\begin{longtable}{lccccccccccc}
\caption{Summary of the 80 pulsars detected in the MWA incoherent search} \label{appendix:80pulsars}\\
\hline
PSR & $P_0$  & DM & $G_{\rm inco}$ & Harm. & S/N & $S_{185}$ & $W_{10}$ & $\alpha_{\rm sc}$ & $\tau_{\rm exp}(\tau_{\rm meas})$ & Fold time & Obs ID \\
& (ms) & (pc cm$^{-3}$) & (K/Jy) &  &  & (mJy)  & (ms / deg) & & (ms) & (s) & \\
\hline
\endfirsthead
\hline
PSR & $P_0$  & DM & $G_{\rm inco}$ & Harm. & S/N & $S_{185}$ & $W_{10}$ & $\alpha_{\rm sc}$ & $\tau_{\rm exp}(\tau_{\rm meas})$ & Fold time & Obs ID \\
& (ms) & (pc cm$^{-3}$) & (K/Jy) &  &  & (mJy)  & (ms / deg) & & (ms) & (s) & \\
\hline
\endhead
\hline
\endfoot
J0034-0534$^{\dag}$	&	1.877	&	13.77	&	0.0129	&	23	&	-	&	-	&	0.6 	/	118.1 	&	-4.0 	&	0.130 	&	2496	&	1137236608\\
J0034-0721	&	943.038	&	10.83	&	0.0137	&	1	&	20	&	218(31)	&	103.1 	/	39.4 	&	-4.3 	&	0.006 	&	2496	&	1137236608\\
J0152-1637$^*$	&	832.817	&	11.91	&	0.0031	&	1	&	15	&	748(103)	&	43.5 	/	18.8 	&	-	&	-	&	1088	&	1137236608\\
J0332+5434$^*$	&	714.468	&	26.78	&	0.0003	&	1	&	12	&	3511(425)	&	11.2 	/	5.6 	&	-4.4 	&	0.043 	&	2304	&	1091309464\\
J0418-4154	&	757.094	&	24.36	&	0.0132	&	1	&	13	&	51(6)	&	22.7 	/	10.8 	&	-	&	-	&	3712	&	1127329112\\
J0437-4715$^{\dag}$	&	5.757	&	2.64	&	0.0224	&	1	&	45	&	536(75)	&	2.6 	/	160.3 	&	-4.0 	&	0.002 	&	2496	&	1123367368\\
J0452-1759$^*$	&	548.978	&	39.9	&	0.0095	&	1	&	12	&	148(18)	&	47.2 	/	30.9 	&	-5.2 	&	1.468 	&	2496	&	1169379256\\
J0520-2553$^{\dag*}$	&	241.654	&	33.77	&	0.0084	&	1/5	&	8	&	97(12)	&	5.7 	/	8.6 	&	-	&	-	&	960	&	1169379256\\
J0534+2200	&	33.7	&	56.77	&	0.008	&	1	&	-	&	-	&	- 	&	-2.9 	&	4.794(7.879) 	&	1152	&	1099414416\\
J0601-0527$^{\dag*}$	&	395.957	&	80.54	&	0.0144	&	1	&	6	&	165(23)	&	29.3 	/	26.6 	&	-2.7 	&	7.235 	&	384	&	1164474312\\
J0630-2834	&	1244.385	&	34.41	&	0.0156	&	1	&	27	&	159(17)	&	126.4 	/	36.6 	&	-3.3 	&	0.007 	&	4800	&	1101491208\\
J0736-6304$^{\dag*}$	&	4862.845	&	18.96	&	0.001	&	1/17	&	8	&	1244(177)	&	683.8 	/	50.6 	&	-	&	-	&	2688	&	1139324488\\
J0742-2822	&	166.764	&	73.73	&	0.0123	&	1	&	12	&	133(15)	&	28.0 	/	60.5 	&	-3.3 	&	6.767(8.967) 	&	3840	&	1101491208\\
J0820-1350	&	1238.205	&	40.91	&	0.0053	&	1/2	&	13	&	286(35)	&	38.6 	/	11.2 	&	-4.2 	&	0.545 	&	960	&	1173793160\\
J0820-4114$^{\dag*}$	&	545.444	&	113.4	&	0.0181	&	1	&	5	&	173(26)	&	- 	&	-4.0 	&	94.684 	&	1216	&	1139239952\\
J0823+0159	&	864.887	&	23.7	&	0.0131	&	1	&	7	&	53(6)	&	40.5 	/	16.9 	&	-4.0 	&	0.019 	&	2176	&	1139324488\\
J0826+2637$^{\dag*}$	&	530.68	&	19.49	&	0.0024	&	1/5	&	11	&	385(47)	&	16.6 	/	11.3 	&	-3.5 	&	0.133 	&	2688	&	1139324488\\
J0835-4510	&	89.39	&	67.77	&	0.0185	&	1	&	-	&	-	&	- 	&	-2.5 	&	49.211(32.049) 	&	1216	&	1139239952\\
J0837+0610	&	1273.8	&	12.83	&	0.0156	&	1	&	22	&	223(27)	&	39.8 	/	11.3 	&	-3.7 	&	0.003 	&	2688	&	1139324488\\
J0837-4135	&	751.619	&	147.2	&	0.0224	&	1	&	168	&	690(83)	&	64.6 	/	30.9 	&	-4.0 	&	13.497(10.514) 	&	1216	&	1139239952\\
J0855-3331$^{\dag}$	&	1267.527	&	86.64	&	0.0185	&	1/21	&	8	&	127(18)	&	99.0 	/	28.1 	&	-4.0 	&	65.400(32.145) 	&	960	&	1139239952\\
J0856-6137$^{\dag*}$	&	962.5	&	95	&	0.0063	&	1/15	&	6	&	167(20)	&	33.4 	/	12.5 	&	-	&	-	&	1984	&	1173020416\\
J0908-1739$^*$	&	401.643	&	15.88	&	0.0159	&	1	&	12	&	54(6)	&	15.7 	/	14.1 	&	-4.0 	&	0.013 	&	2560	&	1173793160\\
J0922+0638	&	430.631	&	27.3	&	0.0067	&	1	&	17	&	238(29)	&	22.8 	/	19.1 	&	-4.4 	&	0.061 	&	2688	&	1139324488\\
J0924-5302$^*$	&	746.333	&	152.9	&	0.0105	&	1/2	&	12	&	172(21)	&	43.1 	/	20.8 	&	-4.0 	&	9.761 	&	3199	&	1173020416\\
J0934-5249$^{\ddag*}$	&	1444.763	&	100	&	0.0116	&	1	&	10	&	95(11)	&	43.7 	/	10.9 	&	-4.0 	&	68.329 	&	3711	&	1173020416\\
J0942-5552$^*$	&	664.38	&	180.16	&	0.0141	&	1/2	&	12	&	118(14)	&	29.2 	/	15.8 	&	-4.0 	&	48.806 	&	2752	&	1173020416\\
J0942-5657$^*$	&	808.153	&	159.74	&	0.0125	&	1/2	&	12	&	107(13)	&	37.9 	/	16.9 	&	-	&	-	&	3711	&	1173020416\\
J0944-1354$^*$	&	570.284	&	12.48	&	0.02	&	1	&	11	&	30(4)	&	13.4 	/	8.4 	&	-	&	-	&	3072	&	1173793160\\
J0946+0951$^{\dag*}$	&	1097.702	&	15.34	&	0.0041	&	1/7	&	12	&	390(47)	&	47.6 	/	15.6 	&	-4.0 	&	0.207 	&	896	&	1139324488\\
J0953+0755	&	253.09	&	2.97	&	0.005	&	1	&	69	&	1130(122)	&	21.3 	/	30.2 	&	-	&	-	&	3968	&	1115381072\\
J0955-5304$^{\dag*}$	&	862.109	&	156.9	&	0.0154	&	3	&	12	&	84(10)	&	27.2 	/	11.4 	&	-4.0 	&	29.284 	&	3392	&	1173020416\\
J0959-4809$^*$	&	670.078	&	92.7	&	0.0128	&	1	&	9	&	134(16)	&	- 	&	-	&	-	&	3711	&	1173020416\\
J1001-5507	&	1436.609	&	130.32	&	0.0154	&	1	&	8	&	113(14)	&	149.7 	/	37.5 	&	-4.0 	&	146.418(133.695) 	&	3711	&	1173020416\\
J1012-2337$^{\dag*}$	&	2517.986	&	22.64	&	0.0091	&	1/2	&	10	&	120(14)	&	118.2 	/	16.9 	&	-	&	-	&	1088	&	1173793160\\
J1057-5226$^*$	&	197.109	&	29.69	&	0.0138	&	1	&	22	&	417(53)	&	- 	&	-4.0 	&	0.012 	&	3711	&	1173020416\\
J1059-5742$^*$	&	1184.965	&	108.7	&	0.0127	&	1	&	11	&	103(12)	&	48.1 	/	14.6 	&	-4.0 	&	19.522 	&	3711	&	1173020416\\
J1112-6926$^{\dag}$	&	820.455	&	148.4	&	0.0107	&	1/13	&	6	&	92(13)	&	51.3 	/	22.5 	&	-	&	-	&	2112	&	1140972392\\
J1116-4122	&	943.19	&	40.53	&	0.0092	&	1	&	23	&	173(21)	&	29.4 	/	11.2 	&	-4.0 	&	4.140 	&	2944	&	1145367872\\
J1136+1551	&	1187.903	&	4.89	&	0.0122	&	1/3	&	21	&	223(31)	&	55.7 	/	16.9 	&	-4.3 	&	0.005 	&	2304	&	1140972392\\
J1141-3107$^{\ddag}$	&	538.451	&	30.77	&	0.0197	&	1/15	&	7	&	34(5)	&	33.7 	/	22.5 	&	-	&	-	&	3584	&	1145367872\\
J1141-6545$^{\dag}$	&	394.085	&	116.08	&	0.0099	&	1/3	&	10	&	207(29)	&	30.8 	/	28.1 	&	-	&	-	&	1152	&	1140972392\\
J1224-6407$^*$	&	216.47	&	97.78	&	0.0074	&	1	&	10	&	160(19)	&	10.1 	/	16.9 	&	-	&	-	&	2496	&	1140972392\\
J1234-3630$^*$	&	569.249	&	59.29	&	0.0172	&	1	&	8	&	42(6)	&	11.6 	/	7.3 	&	-	&	-	&	3584	&	1145367872\\
J1355-5153$^*$	&	644.296	&	112.1	&	0.0138	&	1	&	14	&	166(20)	&	41.2 	/	23.0 	&	-	&	-	&	4864	&	1177428616\\
J1418-3921$^{\ddag*}$	&	1096.792	&	60.49	&	0.0138	&	1/15	&	8	&	103(12)	&	64.6 	/	21.2 	&	-	&	-	&	3392	&	1177428616\\
J1430-6623	&	785.453	&	65.1	&	0.013	&	1	&	11	&	287(35)	&	43.0 	/	19.7 	&	-4.0 	&	0.107 	&	576	&	1131415232\\
J1453-6413	&	179.49	&	71.25	&	0.0115	&	1	&	22	&	907(111)	&	12.6 	/	25.3 	&	-4.0 	&	0.199 	&	576	&	1131415232\\
J1456-6843	&	263.382	&	8.62	&	0.0122	&	1	&	26	&	917(113)	&	10.3 	/	14.1 	&	-3.5 	&	0.003 	&	576	&	1131415232\\
J1507-4352	&	286.75	&	48.7	&	0.0168	&	1	&	9	&	168(24)	&	28.9 	/	36.3 	&	-	&	-	&	2688	&	1177428616\\
J1510-4422$^{\dag*}$	&	943.843	&	84	&	0.0126	&	1/7	&	9	&	115(14)	&	51.2 	/	19.5 	&	-	&	-	&	4480	&	1177428616\\
J1534-5334	&	1368.968	&	24.84	&	0.0173	&	1	&	36	&	329(35)	&	56.1 	/	14.8 	&	-	&	-	&	4864	&	1121173352\\
J1543+0929$^*$	&	748.448	&	34.98	&	0.0118	&	1	&	28	&	278(34)	&	72.3 	/	34.8 	&	-4.5 	&	0.888 	&	1856	&	1178553136\\
J1543-0620$^*$	&	709.056	&	18.28	&	0.0103	&	1	&	12	&	228(28)	&	20.8 	/	10.6 	&	-4.5 	&	0.082 	&	2496	&	1178380936\\
J1607-0032	&	421.809	&	10.69	&	0.0075	&	1	&	12	&	274(33)	&	19.8 	/	16.9 	&	-4.4 	&	0.006 	&	1856	&	1178380936\\
J1645-0317	&	387.699	&	35.76	&	0.021	&	1	&	73	&	798(96)	&	8.0 	/	7.4 	&	-4.4 	&	0.060 	&	704	&	1132030232\\
J1709-1640	&	653.037	&	24.9	&	0.0225	&	1	&	15	&	250(30)	&	25.5 	/	14.1 	&	-6.0 	&	1.150 	&	1152	&	1116090392\\
J1711-5350	&	899.292	&	106.1	&	0.0183	&	1	&	7	&	139(17)	&	52.6 	/	21.1 	&	-	&	-	&	3712	&	1185884600\\
J1731-4744	&	829.982	&	123.06	&	0.0158	&	1	&	37	&	664(80)	&	38.9 	/	16.9 	&	-0.4 	&	8.869 	&	3712	&	1185884600\\
J1751-4657	&	742.405	&	20.43	&	0.0203	&	1	&	29	&	518(63)	&	29.0 	/	14.1 	&	-4.0 	&	0.011 	&	3840	&	1091793216\\
J1752-2806	&	562.58	&	50.37	&	0.0061	&	1	&	55	&	2691(288)	&	16.1 	/	10.3 	&	-4.5 	&	1.105 	&	2944	&	1133329792\\
J1820-0427	&	598.047	&	84.44	&	0.008	&	1	&	9	&	556(82)	&	105.7 	/	63.7 	&	-3.7 	&	14.560(21.484) 	&	3008	&	1116090392\\
J1823-3106	&	284.067	&	50.24	&	0.0154	&	2	&	10	&	307(37)	&	22.2 	/	28.1 	&	-	&	-	&	2944	&	1133329792\\
J1825-0935	&	768.973	&	19.33	&	0.0119	&	1	&	11	&	299(42)	&	36.0 	/	16.9 	&	-3.9 	&	0.127 	&	2624	&	1116090392\\
J1900-2600	&	612.241	&	37.99	&	0.0231	&	1	&	27	&	412(44)	&	45.4 	/	26.7 	&	-2.5 	&	0.468 	&	3712	&	1133329792\\
J1909+1102	&	283.654	&	149.98	&	0.0109	&	1	&	8	&	398(71)	&	58.9 	/	74.8 	&	-3.4 	&	16.572 	&	1984	&	1095506112\\
J1917+1353	&	194.626	&	94.52	&	0.0096	&	1	&	8	&	225(27)	&	16.3 	/	30.1 	&	-2.2 	&	6.152 	&	1599	&	1148063920\\
J1921+2153	&	1337.39	&	12.46	&	0.0079	&	1	&	44	&	834(101)	&	41.8 	/	11.3 	&	-4.9 	&	0.020 	&	3904	&	1095506112\\
J1932+1059	&	226.536	&	3.19	&	0.0116	&	1	&	18	&	394(48)	&	17.7 	/	28.1 	&	-4.2 	&	0.002 	&	2944	&	1095506112\\
J1935+1616	&	358.721	&	158.52	&	0.0103	&	1	&	14	&	305(37)	&	16.9 	/	17.0 	&	-3.4 	&	12.219(7.204) 	&	2560	&	1148063920\\
J1943-1237	&	972.427	&	28.86	&	0.0121	&	1	&	8	&	121(17)	&	15.4 	/	5.7 	&	-4.0 	&	2.462 	&	1792	&	1152636328\\
J2018+2839	&	557.98	&	14.2	&	0.0044	&	1	&	18	&	878(106)	&	26.2 	/	16.9 	&	-4.7 	&	0.043 	&	2688	&	1095506112\\
J2022+2854$^{\dag}$	&	343.427	&	24.63	&	0.0074	&	1/11	&	7	&	450(62)	&	6.7 	/	7.0 	&	-4.6 	&	0.022 	&	384	&	1131957328\\
J2046-0421	&	1546.891	&	35.86	&	0.011	&	1	&	10	&	118(16)	&	68.5 	/	15.9 	&	-4.0 	&	0.077 	&	3968	&	1152636328\\
J2048-1616	&	1961.768	&	11.48	&	0.0091	&	1	&	16	&	331(37)	&	114.9 	/	21.1 	&	-2.9 	&	0.002 	&	2816	&	1097404000\\
J2145-0750$^{\dag}$	&	16.05	&	9	&	0.0142	&	5	&	26	&	307(39)	&	- 	&	-3.1 	&	0.003 	&	4863	&	1118168248\\
J2219+4754	&	538.485	&	43.5	&	0.0012	&	1	&	23	&	1108(133)	&	16.8 	/	11.3 	&	-4.2 	&	0.186 	&	5056	&	1129464688\\
J2241-5236$^{\dag}$	&	2.187	&	11.41	&	0.0188	&	17	&	19	&	136(17)	&	0.3 	/	47.8 	&	-	&	-	&	5056	&	1129464688\\
J2256-1024$^{\dag*}$	&	2.295	&	13.78	&	0.0134	&	11	&	11	&	137(29)	&	0.5 	/	80.5 	&	-	&	-	&	4864	&	1133775752\\
J2330-2005$^*$	&	1643.784	&	8.55	&	0.0028	&	1/21	&	7	&	341(47)	&	77.0 	/	16.9 	&	-2.3 	&	0.002 	&	1152	&	1133775752\\
\hline
\end{longtable}
\vspace{0mm}
\noindent
\textbf{Notes.} $^{\dag}$ Detected via harmonic matching with the ATNF catalog. \\
$^{\ddag}$ Recovered using fixed downsampling (factor 8). \\
$^*$ First detections at 185 MHz.

\begin{table*}
\centering
\caption{Broadband morphological spectral fitting results of 75 pulsars by \texttt{pulsar\_spectra} toolkit}
\label{appendix:all_models_final_results}
\setlength{\tabcolsep}{6pt}
\renewcommand{\arraystretch}{1.12}

\begin{tabular}{lc lccc lcc}
\toprule
\multicolumn{2}{c}{\textbf{Simple Power Law}} &
\multicolumn{3}{c}{\textbf{High-Frequency Cutoff Power Law}} &
\multicolumn{4}{c}{\textbf{Broken Power Law}}
 \\
\cmidrule(lr){1-2}\cmidrule(lr){3-5}\cmidrule(l){6-9}
\textbf{PSR} & $\alpha$ &
\textbf{PSR} & $\alpha$ & $\nu_{\rm c}$ (MHz) & \textbf{PSR} & $\alpha_1$ & $\alpha_2$ & $\nu_{\rm b}$ (MHz)\\
\midrule
J0418$-$4154 & $-1.91 \pm 0.15$ & J0823$+$0159 & $-1.11 \pm 0.14$ & $5110 \pm 358$ & J0452$-$1759 & $-0.36 \pm 0.08$ & $-2.13 \pm 0.11$ & $702.00 \pm 65.70$  \\ 
J0520$-$2553 & $-1.24 \pm 0.36$ & J0955$-$5304 & $-1.01 \pm 0.53$ & $1570 \pm 71$ & J0742$-$2822 & $-0.89 \pm 0.05$ & $-1.73 \pm 0.11$ & $606.00 \pm 0.39$ \\ 
J0856$-$6137 & $-2.20 \pm 0.08$ & J1116$-$4122 & $-1.38 \pm 0.08$ & $3450 \pm 215$ & J0820$-$4114 & $-0.59 \pm 0.27$ & $-2.04 \pm 0.26$ & $376.00 \pm 103.00$ \\ 
J0924$-$5302 & $-2.00 \pm 0.07$ & J1709$-$1640 & $-0.91 \pm 0.03$ & $24800\pm 2750$ & J0855$-$3331 & $-1.70 \pm 0.11$ & $-2.67 \pm 0.28$ & $950.00 \pm 0.37$ \\
J0934$-$5249 & $-2.15 \pm 0.08$ & & & & J1001$-$5507 & $-0.65 \pm 0.59$ & $-2.59 \pm 0.28$ & $633.00 \pm 263.00$ \\ \cmidrule(l){6-9}
J0942$-$5657 & $-2.23 \pm 0.08$ & & & &\multicolumn{4}{c}{\textbf{Low-Frequency Turnover Power Law}} \\ \cmidrule(l){6-9}
J0944$-$1354 & $-1.90 \pm 0.09$ & & & &\textbf{PSR} & $\alpha$ & $\beta$ & $\nu_{\rm peak}$ (MHz) \\ \cmidrule(l){6-9}
J0959$-$4809 & $-1.59 \pm 0.09$ & & & & J0152$-$1637 & $-1.94 \pm 0.44$ & $0.99 \pm 0.70$ & $48.20 \pm 4.13$ \\
J1012$-$2337 & $-2.62 \pm 0.09$ & & & & J0332$+$5434 & $-2.09 \pm 0.13$ & $1.36 \pm 0.22$ & $157.00 \pm 5.06$ \\
J1112$-$6926 & $-1.77 \pm 0.17$ & & & & J0601$-$0527 & $-1.78 \pm 0.07$ & $2.10 \pm 1.49$ & $183.00 \pm 15.70$ \\
J1141$-$3107 & $-1.57 \pm 0.42$ & & & & J0820$-$1350 & $-3.90 \pm 0.53$ & $0.27 \pm 0.02$ & $65.70 \pm 18.80$ \\
J1224$-$6407 & $-1.20 \pm 0.12$ & & & & J0826$+$2637 & $-1.87 \pm 0.23$ & $0.55 \pm 0.19$ & $59.00 \pm 8.04$ \\
J1418$-$3921 & $-1.98 \pm 0.06$ & & & & J0837$+$0610 & $-6.69 \pm 0.56$ & $0.22 \pm 0.01$ & $68.40 \pm 6.81$ \\
J1507$-$4352 & $-2.38 \pm 0.13$ & & & & J0837$-$4135 & $-2.44 \pm 0.21$ & $1.04 \pm 0.21$ & $271.00 \pm 11.40$ \\
J1510$-$4422 & $-1.42 \pm 0.54$ & & & & J0908$-$1739 & $-1.52 \pm 0.07$ & $2.10 \pm 0.13$ & $119.00 \pm 2.08$ \\
J1711$-$5350 & $-2.50 \pm 0.10$ & & & & J0922$+$0638 & $-1.90 \pm 0.22$ & $0.84 \pm 0.30$ & $45.90 \pm 1.15$ \\
J1823$-$3106 & $-1.58 \pm 0.06$ & & & & J0942$-$5552 & $-2.97 \pm 0.05$ & $1.06 \pm 0.03$ & $298.00 \pm 8.87$ \\
J2241$-$5236 & $-2.06 \pm 0.03$ & & & & J0953$+$0755 & $-3.56 \pm 2.96$ & $0.29 \pm 0.42$ & $123.00 \pm 44.90$ \\
J2256$-$1024 & $-2.69 \pm 0.46$ & & & & J1057$-$5226 & $-2.88 \pm 0.41$ & $2.10 \pm 0.40$ & $195.00 \pm 18.40$ \\
\addlinespace
\cmidrule(lr){1-5}
\multicolumn{5}{c}{\textbf{Double Turn-over Spectrum}}& J1059$-$5742 & $-2.53 \pm 0.28$ & $2.10 \pm 1.68$ & $193.00 \pm 21.50$ \\
\cmidrule(lr){1-5}
\textbf{PSR} & $\alpha$ & $\beta$ & $\nu_{\rm peak}$ (MHz) & $\nu_{\rm c}$ (MHz) & J1141$-$6545 & $-2.73 \pm 0.28$ & $2.10 \pm 1.03$ & $224.00 \pm 14.00$ \\
\cmidrule(lr){1-5}
J0034$-$0721 & $-3.34 \pm 0.18$ & $0.78 \pm 0.06$ & $73.10 \pm 1.97$ & $14300 \pm 9580$ & J1355$-$5153 & $-3.10 \pm 0.13$ & $2.10 \pm 1.54$ & $173.00 \pm 9.00$ \\
J0437$-$4715 & $-1.92 \pm 0.0$ & $0.30 \pm 0.00$ & $50.90 \pm 0.02$ & $17800 \pm 65$ & J1430$-$6623 & $-1.94 \pm 0.18$ & $2.10 \pm 1.31$ & $210.00 \pm 19.00$ \\
J0630$-$2834 & $-1.86 \pm 0.03$ & $2.10 \pm 0.12$ & $83.50 \pm 0.39$ & $105000 \pm 61300$ & J1453$-$6413 & $-2.67 \pm 0.17$ & $2.10 \pm 0.13$ & $194.00 \pm 8.82$ \\
J0946+0951 & $-5.38 \pm 2.18$ & $0.73 \pm 0.45$ & $52.60 \pm 3.05$ & $14200 \pm 6460$ & J1456$-$6843 & $-1.94 \pm 0.08$ & $2.10 \pm 1.43$ & $219.00 \pm 2.78$ \\
J1136+1551 & $-1.79 \pm 0.07$ & $1.12 \pm 0.18$ & $67.20 \pm 3.44$ & $227000 \pm 102000$ & J1534$-$5334 & $-2.66 \pm 0.37$ & $2.10 \pm 0.26$ & $194.00 \pm 25.90$ \\
J1543+0929 & $-1.92 \pm 0.17$ & $1.48 \pm 0.32$ & $82.80 \pm 2.91$ & $5520 \pm 312$ & J1645$-$0317 & $-2.36 \pm 0.09$ & $0.80 \pm 0.05$ & $145.00 \pm 1.07$ \\
J1543$-$0620 & $-2.11 \pm 0.18$ & $1.49 \pm 0.30$ & $80.20 \pm 3.64$ & $47500 \pm 25300$ & J1731$-$4744 & $-2.09 \pm 0.08$ & $2.10 \pm 0.29$ & $215.00 \pm 2.48$ \\
J1607$-$0032 & $-1.93 \pm 0.14$ & $1.62 \pm 0.26$ & $80.00 \pm 3.52$ & $105000 \pm 53500$ & J1751$-$4657 & $-2.79 \pm 0.18$ & $2.10 \pm 1.43$ & $185.00 \pm 14.70$ \\
J1752$-$2806 & $-3.01 \pm 2.02$ & $1.07 \pm 1.48$ & $152.00 \pm 20.40$ & $227000 \pm 104000$ & J1900$-$2600 & $-2.31 \pm 0.22$ & $1.01 \pm 0.23$ & $128.00 \pm 1.84$ \\
J1820$-$0427 & $-2.39 \pm 0.03$ & $2.10 \pm 0.04$ & $170.00 \pm 2.26$ & $49200 \pm 43000$ & J1909$+$1102 & $-2.63 \pm 0.11$ & $2.10 \pm 1.32$ & $125.00 \pm 9.93$ \\
J1825$-$0935 & $-0.99 \pm 0.09$ & $2.10 \pm 0.39$ & $55.70 \pm 6.21$ & $12400 \pm 1220$ & J1917$+$1353 & $-2.61 \pm 1.02$ & $0.99 \pm 1.37$ & $120.00 \pm 107.00$ \\
J1932+1059 & $-1.08 \pm 0.033$ & $1.4 \pm 0.26$ & $67.40 \pm 6.25$ & $51500 \pm 6210$ & J1921$+$2153 & $-2.66 \pm 0.03$ & $2.10 \pm 0.28$ & $79.10 \pm 2.64$ \\
& & & & & J1935$+$1616 & $-3.98 \pm 1.51$ & $0.43 \pm 0.27$ & $298.00 \pm 12.80$ \\
& & & & & J1943$-$1237 & $-2.21 \pm 0.16$ & $2.10 \pm 1.89$ & $155.00 \pm 18.00$ \\ 
& & & & & J2018$+$2839 & $-8.00 \pm 0.56$ & $0.13 \pm 0.00$ & $151.00 \pm 4.55$ \\
& & & & & J2022$+$2854 & $-4.00 \pm 0.75$ & $0.59 \pm 0.09$ & $187.00 \pm 23.10$ \\ 
& & & & & J2046$-$0421 & $-3.43 \pm 0.38$ & $0.57 \pm 0.07$ & $170.00 \pm 22.10$ \\
& & & & & J2048$-$1616 & $-4.00 \pm 0.50$ & $0.15 \pm 0.01$ & $49.40 \pm 5.36$ \\
& & & & & J2145$-$0750 & $-1.72 \pm 0.07$ & $1.00 \pm 0.11$ & $45.10 \pm 0.31$ \\
& & & & & J2219$+$4754 & $-2.93 \pm 0.32$ & $1.10 \pm 0.32$ & $72.10 \pm 5.44$ \\
& & & & & J2330$-$2005 & $-2.05 \pm 0.07$ & $2.10 \pm 1.26$ & $75.70 \pm 0.77$ \\
\bottomrule
\end{tabular}
\raggedright \textbf{Note.} Definition of the models and parameters are by \citet{Jankowski2018} and \citet{Swainston2022}.
\end{table*}

\begin{figure*}
    \centering
    \includegraphics[width=\textwidth]{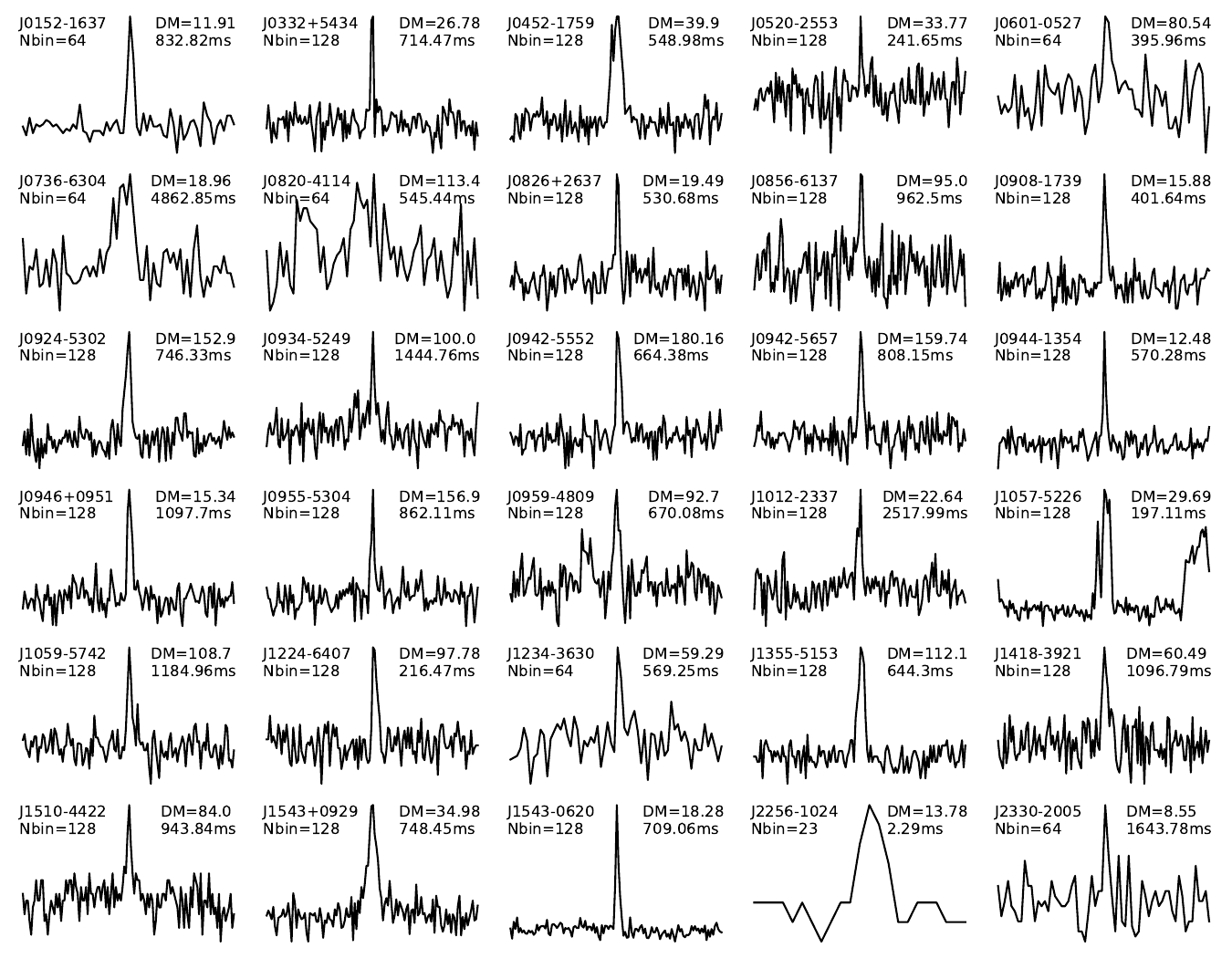}
    \caption{The integrated pulse profiles of 30 pulsars searched from our 48 MWA-VCS incoherently summed data, reported at 185 MHz for the first time.}
    \label{fig:30pulsars}
\end{figure*}






\bsp	
\label{lastpage}
\end{document}